\documentclass[12pt]{elsart}
\pdfoutput=1
\usepackage{graphicx}        
\usepackage{natbib}         
\usepackage{amssymb}        
\usepackage{color}           
\usepackage{url}             
\usepackage{courier}
\usepackage{sidecap}
\usepackage{lscape}
\usepackage{color}
\newlength\abovecaptionskip
\newlength\belowcaptionskip
\newcommand{\figAIA}{7}   

\newcommand{\eg}{\textit{e.g.}}

\renewcommand{\vec}[1]{ {\mathbf #1} }

\newcommand{\bb}{ \vec B}
\newcommand{\jj}{ \vec j}

\begin{document}
\begin{frontmatter}
 
\title{Topological Analysis of Emerging Bipole Clusters Producing Violent Solar Events}

\author[iafe,facu]{C.H. Mandrini\corauthref{cor}\thanksref{coni}}
\corauth[cor]{Corresponding author, e-mail: mandrini@iafe.uba.ar}
\author[Meudon]{B. Schmieder}
\author[Meudon]{P. D\'emoulin}
\author[Nanjing]{Y. Guo}
\author[iafe,facu]{G.D. Cristiani\thanksref{coni}}

\thanks[coni]{Member of the Carrera del Investigador Cient\'\i fico, 
        CONICET, Argentina}

\address[iafe]{Instituto de Astronom\'\i a y F\'\i sica del Espacio (IAFE), CONICET-UBA, Buenos Aires, Argentina}
\address[facu]{Facultad de Ciencias Exactas y Naturales (FCEN), UBA, Buenos Aires, Argentina}
\address[Meudon]{Observatoire de Paris, Meudon 92190, France}             
\address[Nanjing]{School of Astronomy and Space Science, Nanjing University, Nanjing 210093, China}

\begin{abstract}
During the rising phase of Solar Cycle 24 tremendous activity occurred on the Sun with fast and compact emergence of magnetic flux leading to bursts of flares (C to M and even X-class). We investigate the violent events occurring in the cluster of two active regions (ARs), NOAA numbers 11121 and 11123, observed in November 2010 with instruments onboard the {\it Solar Dynamics Observatory} and from Earth. Within one day the total magnetic flux increased by 70\,\% with the emergence of new groups of bipoles in AR 11123. 
From all the events on 11 November, we study, in particular, the ones starting at around 07:16 UT in GOES soft X-ray data and the brightenings preceding them. A magnetic-field topological analysis indicates the presence of null points, associated separatrices and quasi-separatrix layers (QSLs) where magnetic reconnection is prone to occur. The presence of null points is confirmed by a linear and a non-linear force-free magnetic-field model. Their locations and general characteristics are similar in both modelling approaches, which supports their robustness. However, in order to explain the full extension of the analysed event brightenings, which are not restricted to the photospheric traces of the null separatrices, we compute the locations of QSLs.   
Based on this more complete topological analysis, we propose a scenario to explain the origin of a low-energy event preceding a filament eruption, which is accompanied by a two-ribbon flare, and a consecutive confined flare in AR 11123. The results of our topology computation can also explain the locations of flare ribbons in two other events, one preceding and one following the ones at 07:16 UT.  Finally, this study provides further examples where flare-ribbon locations can be explained when compared to QSLs and only, partially, when using separatrices.  
\end{abstract}

\begin{keyword}
Sun: flares \sep Sun: filament eruptions \sep Sun: magnetic fields
\end{keyword}
\end{frontmatter}

\section{Introduction}
\label{S_Introduction}
 
Active regions (ARs) consist of strong concentrations of magnetic flux that are continuously evolving.
Observations of the evolution of an AR demonstrate that there are usually several episodes of new flux emergence during the AR lifetime, including the appearance of new bipoles, or even new sunspots, in mature and apparently stable ARs \citep[\it e.g.][]{Zuccarello08,Valori12}. 


The first signature of flux emergence in the photosphere is the alignment of granules and horizontal diverging flows 
\citep{Spruit81,Toriumi12}. Rising motions of subsurface magnetic fields are also detectable by helioseismological analysis \citep{Toriumi13}. Later, pores or faculae appear at the solar surface \citep{Strous96}. Direct emergence through the photosphere of a flux tube forming a filament has been proposed to interpret observations of the vector magnetic field in a filament channel \citep{Okamoto09,Lites10}. In the chromosphere and transition region we see a hierarchy of loops, including the formation of arch-filament systems with upward plasma motion at the loop tops and downward ones along their legs \citep{Malherbe98,Mandrini02,Schmieder04,Romano07,Zuccarello08}.
Low-altitude magnetic reconnection has also been observed as chromospheric brightenings at the footpoints of new flux during its emergence \citep{Pariat04,Pariat07,Guglielmino10}. Other byproducts of flux emergence can be jets, plasmoid ejections \citep[see {\it e.g.} the example of a dynamic blob ejection in][]{Srivastava12,Kumar13-1} and large-scale transient phenomena such as flares and CMEs \citep[see {\it e.g.}][]{Schrijver09,Kumar13-2}.

	A common feature during magnetic-flux emergence is the presence of magnetic tongues in the two main polarities at both sides of the polarity inversion line \citep[PIL: ][]{Lopez-Fuentes00,Chandra09,Luoni11,Poisson12}. 
Furthermore, as the magnetic field emerges at the photospheric level, magnetic elements of opposite polarity typically show shearing motions of a few km s$^{-1}$ along the PIL \citep[\eg\ ][]{Strous96}. This evolution and, in particular, magnetic tongues are evidence that a twisted magnetic-flux tube is progressively emerging, as shown in numerical simulations 
\citep[\eg\ ][ and references therein]{Hood09,MacTaggart10}. 

Coronal magnetic-field modelling, or photospheric-field extrapolation, is commonly done with various levels of refinement.  Linear force-free field (LFFF) models [$\jj = \alpha \bb$, with $\alpha$ constant] have been applied with success in pre-flaring and flaring ARs to interpret and understand the observed events \citep[\eg\ ][]{Mandrini06,Luoni07,Cristiani07,Chandra11,Reid12}. Non-linear force-free field (NLFFF) models 
[$\jj = \alpha \bb$, $\alpha$ variable in space] are more sophisticated extrapolation methods. At present, this is an active area of research because photospheric vector magnetograms do not fully provide the needed boundary conditions and, therefore, the adopted assumption(s) largely determine the solution \citep[\eg\ ][and references therein]{DeRosa09,Valori10}. 

A typical application of magnetic-field extrapolations is the computation of the coronal-field topology.
When magnetic null points are found in the corona, the magnetic-field configuration in their surroundings typically shows a fan and spine structure \citep{Longcope05b,Pontin11}. Fan and spine reconnection solutions around nulls have been obtained both numerically \citep{Craig96,Craig99,Wyper13} and analytically \citep{Ji01,WilmotSmith11}.  Some flares show evidence of the existence of magnetic null points in their reconstructed coronal magnetic configurations \citep[\eg\ ][]{Mandrini91,Mandrini93,Parnell94,Aulanier00,Manoharan05,Luoni07,Reid12}, but not all of them \citep[\eg\ ][]{Demoulin94b,Mandrini96,Bagala00,Schmieder07,Savcheva12}.

To understand the origin and evolution of flares in magnetic configurations without null points,  \cite{Demoulin96a} generalised the concept of separatrices by introducing quasi-separatrix layers (QSLs), which are 3D thin volumes where the field-line linkage experiences a drastic change. QSLs can be found in cases where no magnetic null is present, but 
when a null exists, a separatrix is located at the core of the QSL related to this null \citep{Masson09}.
QSLs are preferred sites for the buildup of current layers and, therefore, as with separatrices, the sites where magnetic field reconnection naturally occurs as shown in several numerical examples
\citep{Milano99,Aulanier05,Buchner06,Pariat06b,WilmotSmith09a,Effenberger11,Janvier13}. The computation of QSLs has allowed us to understand the location of flare ribbons and of energy release in flares
\citep[\eg\ ][]{Demoulin97,Bagala00,Mandrini06,Cristiani07,Savcheva12,Janvier13}.  Moreover, the properties of QSLs depend weakly on the details of the magnetic-field model and, therefore, QSLs are a very robust tool to learn about the characteristics of flare energy release \citep[see the reviews by ][]{Longcope05b,Demoulin06,Mandrini10}. 

In this article we describe the activity and analyse the topology of  a set of two active regions (ARs), NOAA numbers 11121 and 11123, observed during the rising phase of Solar Cycle 24. 
The article is organised as follows. In Section~\ref{S_Observations} we present the data used in our analysis.
In Section~\ref{S_Overview} we summarise the evolution of the AR complex, describe several episodes of flux emergence, give an overview of a series of flares in AR 11123 
and describe in detail the events starting around 07:16 UT in GOES soft X-ray data on 11 November 2010.
Section~\ref{S_Model_Topo_Null} presents a large-scale LFFF model of both ARs and a local NLFFF model of AR 11123. We find the presence of magnetic null points with similar characteristics in both magnetic-field models. The relationship between magnetic-field lines in the null neighbourhoods and the location of a low-energy event, preceding a filament eruption, and the ribbons of a confined flare following the eruption,  
provides a first hint to the interpretation and understanding of these events. However, the field lines computed from the nulls are not sufficient to explain the spatial extension of the emission. Then we summarise QSL properties in Section~\ref{S_QSLs_General}, while we analyse the connectivity of field lines issued from QSLs and their association to the observed brightenings to explain this series of events in Sections~\ref{S_QSLs_Fil-erup} and ~\ref{S_QSLs_FL2}.
In Section~\ref{S_QSLs_Other}, we show the observations of two other events in AR 11123 that could be explained by the results of the same topological computation. Finally, in Section~\ref{S_Interpretation}, we
discuss and summarise our results.

\section{Observations}
\label{S_Observations}

The {\it Atmospheric Imaging Assembly} \citep[AIA: ][]{Lemenetal:2012}, onboard the \textit{Solar Dynamics Observatory} \citep[SDO: ][]{Pesnell12}, takes images of the Sun in seven extreme-ultraviolet (EUV) and three UV-visible wavelength bands with a rapid cadence of 12 seconds and a short exposure time of 0.12 seconds. The pixel sampling of SDO/AIA is $0.6''$ and its field of view extends to 1.3 ${\rm R_{\odot}}$. We mainly use the images obtained with the following three EUV filters: 171 \AA (log~$T \approx$ 5.9), 211 \AA (log~$T \approx$ 6.2, 7.2), and 304 \AA (log~$T \approx$ 4.8). SDO/AIA data are analysed using standard codes in the Solar Software package.

The {\it Helioseismic and Magnetic Imager} \citep[HMI: ][]{Schou12}, onboard SDO, observes the full disk of the Sun with
two 4k$\times$4k CCDs. It provides two sets of magnetic-field measurements simultaneously, {\it i.e.} the line-of-sight field at a cadence of 45 seconds with a precision of 10 G and the vector magnetic field with a lower cadence. The pixel sampling is $0.5''$. SDO/HMI is a filtergraph, which takes an image at a wavelength band and scans that wavelength to sample the spectral line. To obtain the vector magnetic field, SDO/HMI obtains raw filtergrams at six different wavelengths and four to six polarisation states (depending on the polarisation modulation) in the Fe {\sc i} 6173 \AA\ spectral line. The four Stokes parameters [$I, Q, U$ and $V$] at the six wavelengths are computed from the raw data after  
calibration. Then all of the Stokes parameters are averaged over 12 minutes to increase the signal-to-noise ratio, to filter the $p$-modes and to lower the cadence (to limit the data volume). The vector magnetic-field components are computed using the code of Very Fast Inversion of the Stokes Vector \citep[{\sf VFISV}:][]{Borrero11}. 
{\sf VFISV} adopts a 
line-synthesised model that is based on the Milne--Eddington atmosphere. The code has been optimised for HMI data preprocessing for which the damping constant has been set to 0.5 and the filling factor to 1. However, for users who do not want to do the Stokes inversion themselves, the magnetic-field data are available at the Joint Science Operations Center (\url{http://jsoc.stanford.edu/ajax/lookdata.html}).

The {\it T\'elescope H\'eliographique pour l'Etude du Magn\'etisme et des Instabilit\'es Solaires} \citep[THEMIS: ][]{Molodij96}, in the Canary Islands, obtains the full Stokes profiles of the Fe {\sc i} lines, 6302.5 and 6301.5 \AA, with an exposure time of 300 ms. Unlike HMI, THEMIS is a spectrograph with a slit, which has a high dispersion of 12.5 m\AA\ per pixel but needs to scan the solar surface. The four Stokes profiles [$I, Q, U$ and $V$] are computed from the raw spectra after calibration \citep{Bommier02}. The vector magnetic field is finally obtained by the inversion code {\sf UNNOFIT} \citep{Bommier07}, which is also based on the Milne--Eddington atmosphere assumption. 

\begin{landscape}
\begin{figure} 
\centering
\includegraphics[width=1.4\textwidth]{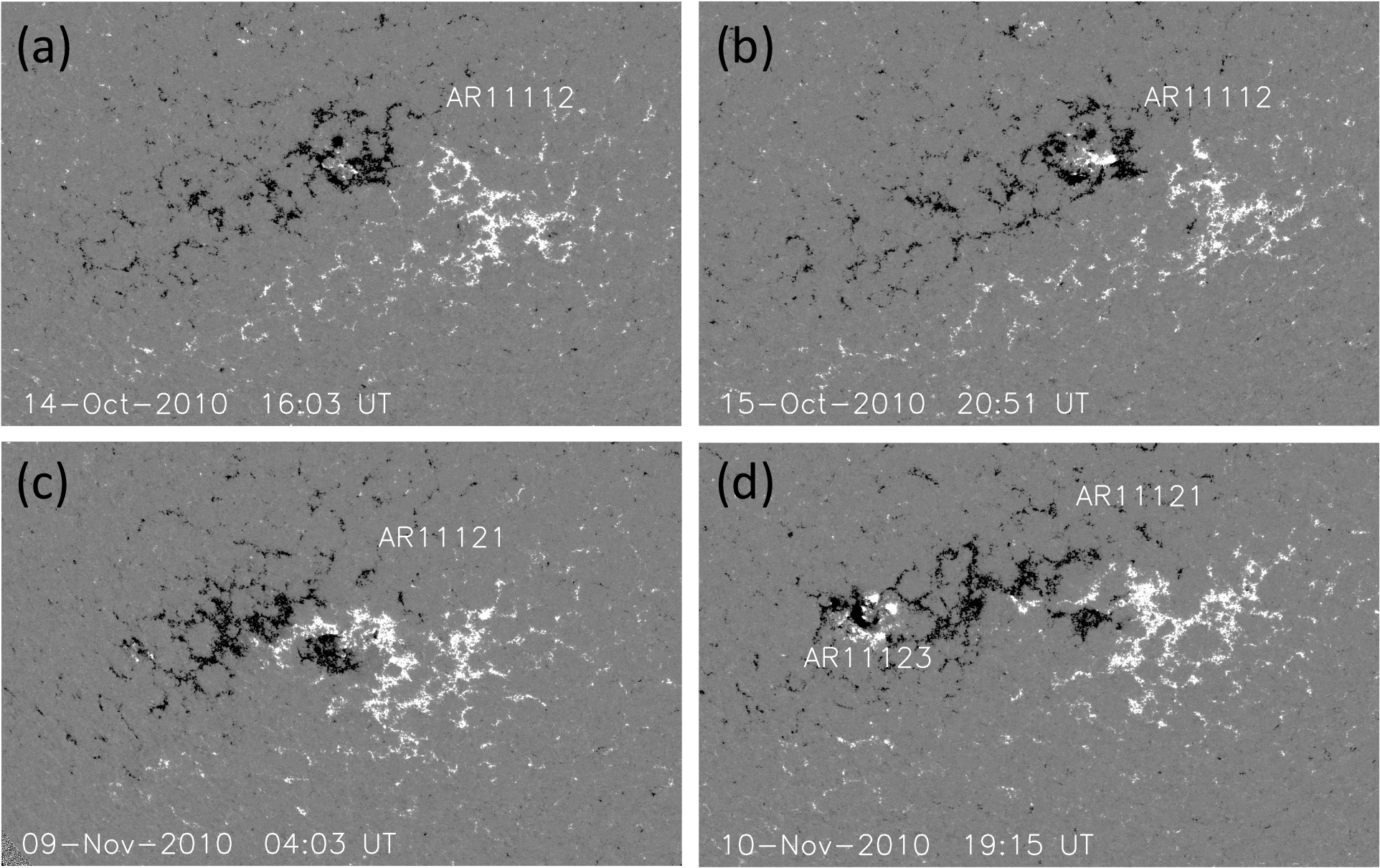}
\caption{Large magnetic-field complex in the southern hemisphere during (a,b) October and (c,d) November 2010. The magnetograms show some of the bipole emergences within a long-living AR.  
AR 11123 was born in the negative-field environment of the following polarity of AR 11121 during November rotation (panel (d)). White (black) regions are positive (negative) line-of-sight magnetic polarities. The magnetic-field values have been saturated above (below) 500 G (-500 G). The size of each panel is $800''$ in the E--W (East--West) and $500''$ in the N--S (North--South) direction. The centre of each panel in heliographic coordinates changes from one solar rotation to the other, being located at [-80$''$,-455$''$] in the first panel; Sun's centre is at [0$''$,0$''$].}.   
\label{HMI_long-term}
\end{figure}
\end{landscape}

\begin{landscape}
\begin{figure} 
\centering
\includegraphics[width=1.62\textwidth]{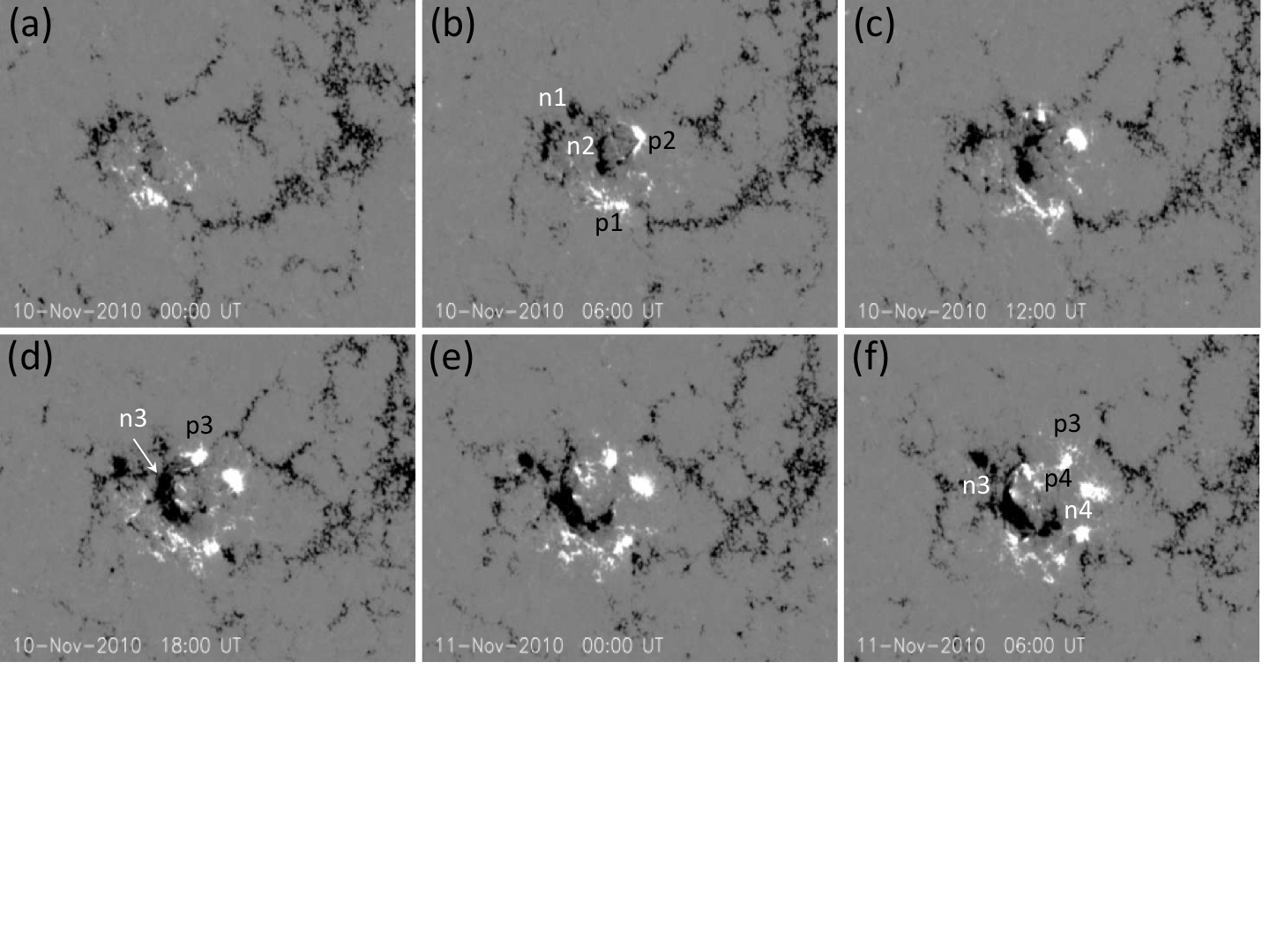}
\vspace*{-4.7cm}
\caption{(a--f) Emergence of different bipoles forming the new AR 11123 between 10 and 11 November 2010. White (black) regions are positive (negative) line-of-sight magnetic polarities. The magnetic-field values have been saturated above (below) 500 G (-500 G). The size of each panel is $240''$ ($150''$) in the horizontal (vertical) direction. The centre of each panel in heliographic coordinates changes with solar rotation, being located at [-415$''$,-413$''$] in the first panel; Sun's centre is at [0$''$,0$''$]. Labels in the vicinity of the polarities in (b), (d) and (f) indicate the different emerging bipoles of AR 11123 (see Section~\ref{S_Emergence23}).}
\label{EMF}
\end{figure}
\end{landscape}

\section{Nested AR Complex}
\label{S_Overview}

\subsection{Long-term Magnetic Evolution and New Flux Emergence}
\label{S_Evolution}

Solar Cycle 24 started rising at the beginning of 2010, after a long solar minimum that lasted nearly two years. Between the beginning of 2010 and the end of 2011, large long-living ARs were observed (see an example in Figure~\ref{HMI_long-term}).
In general, they displayed strong activity with flares, filament eruptions and coronal mass ejections (CMEs) \citep{Schrijver11,Liu12b}. Our present study is focussed on a complex formed by two ARs located in the southern solar hemisphere during November 2010.
This complex was formed by the long-living preceding AR 11121 and the new emerging trailing AR 11123 (Figure~\ref{HMI_long-term}(d)).  The preceding AR was born on the far side of the Sun and appeared on the solar disc in July 2010 as AR 11089. It was a mainly bipolar region, although significant flux emergence occurred all through its lifetime. By August 2010, it was observed decaying as AR 11100; although again, due to new flux emergence both in its preceding and following polarities, it reappeared as AR 11106 in September 2010 (see \cite{Guo13} for the analysis of the activity associated with a small emerging bipole observed during this rotation and \citet{Schmieder13b} for the study of a jet and associated topology). By October 2010, the AR (AR 11112) was very extended and in the decaying phase; although again new flux emergence occurred in its trailing main polarity (Figure~\ref{HMI_long-term}(a), (b)). In November 2010, the remnant main polarities appeared only as facular regions, named AR 11121, accompanied by the emergence of a reverse bipole as the AR crossed the eastern limb (Figure~\ref{HMI_long-term}(c)). During this rotation, a series of smaller bipoles emerged violently within the following polarity and formed AR 11123 (Figure~\ref{HMI_long-term}(d)), which we analyse in detail in this article. 
 
A detailed analysis of the evolution of the spatial distribution \citep[presence of tongues; ][]{Luoni11} of the polarities during the first AR rotation indicates that its magnetic helicity sign was positive (Poisson {\it et al.} 2013, private communication).  This is also evident from the shape of the emerging bipoles shown in Figure~\ref{HMI_long-term} during the October and November rotations. 
In particular, magnetic tongues were also present during the first stages of the appearance of a bipole [n2--p2] in AR 11123 (see Figure~\ref{EMF}(b)). All of these clues indicate the emergence of twisted flux tubes with positive magnetic helicity.

\begin{figure} 
\centerline{\includegraphics[width=0.9\textwidth]{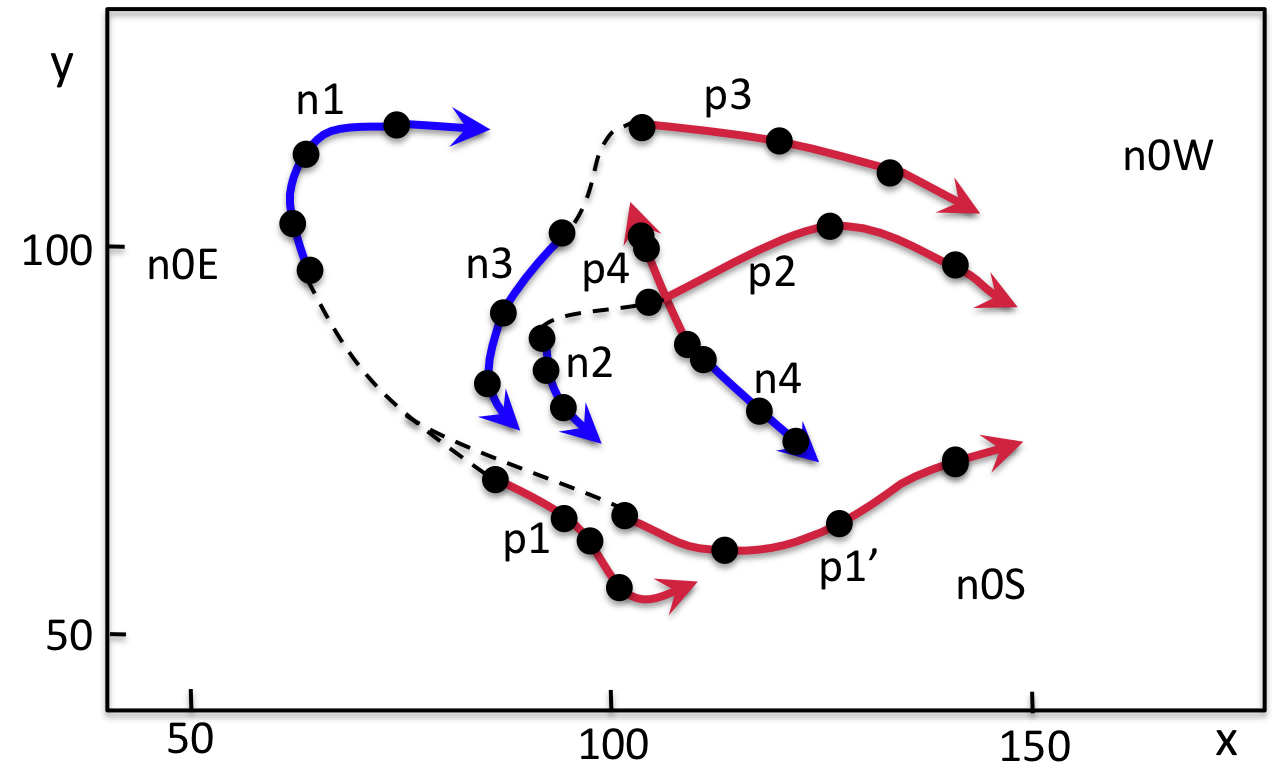}}
\caption{Summary of the evolution of the polarities during the emergence of AR 11123. The polarities are labelled as in Figure~\ref{EMF}. Magenta (blue) colour is used for the path of positive (negative) polarities. The head of the arrows indicate the polarity positions at 12:00 UT on 11 November 2010. Neighbour dots and corresponding arrow head are separated backwards by 12 hours. An extension of the arrow head indicates that a polarity was present at $\approx$ 00:00 UT on 10 November 2010 (initial time of 
{\sf magnetic-evolution.mpg}). Dashed lines link the polarities belonging to the same bipole when they appear separated.}
\label{evolution}
\end{figure}

\begin{SCfigure} 
\vspace{2cm}
\centering
\includegraphics[width=0.72\textwidth,clip=]{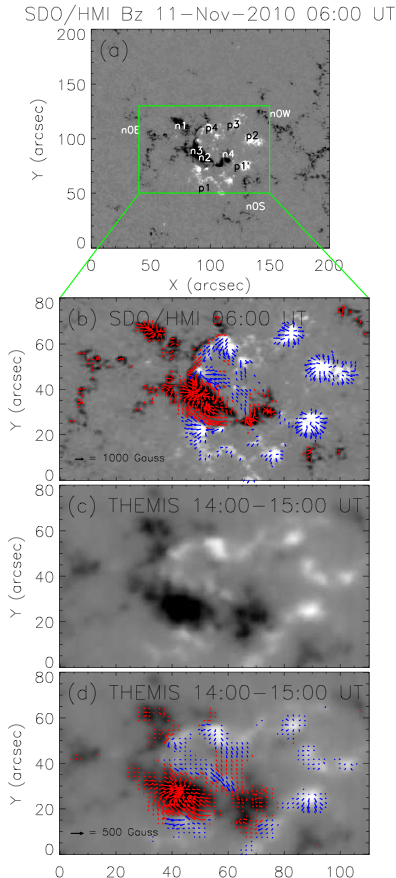}
\caption{(a) and (c) Vertical [$B_z$] magnetic-field component observed with HMI at 06:00 UT and with THEMIS between 14:00--15:00 UT, respectively. (b) and (d) Vertical and horizontal components observed with both instruments. White (black) regions correspond to positive (negative) polarities. The vertical-field values have been saturated above (below) 1000 G (-1000 G) for HMI and above (below) 500 G (-500 G) for THEMIS. The blue (red) arrows indicate the direction of the horizontal-field component for the positive (negative) vertical-field regions. The arrow density is uniform above (below) a vertical-field value of 250 G (-250 G) for HMI and 40 G (-40 G) for THEMIS. The length of the arrows is proportional to the horizontal-field intensity. Labels 
on the polarities in panel (a) indicate the different emerging bipoles (see Section~\ref{S_Emergence23}).}
\label{themis}
\end{SCfigure}

\subsection{Emergence of AR 11123}
\label{S_Emergence23}

Several bipoles emerged successively between 9 and 10 November, forming a cluster within the trailing negative polarity of AR 11121. The different emergence episodes are shown in Figure~\ref{EMF} and all of the bipoles are labelled at a later time in Figure~\ref{themis}(a). We indicate the pre-existing field with labels n0W for the western, n0S for the southern and n0E for the eastern zones. The emergences lead to the birth of AR 11123. The magnetic-field evolution is shown in the movie {\sf magnetic-evolution.mpg}, attached as online material, in which the white square surrounds AR 11123. 
The global evolution of the main polarities is shown in Figure~\ref{evolution}. It shows that the emergence of AR 11121 was far from the classical divergence observed during the emergence of bipolar ARs. 
The first bipole, with polarities labelled as n1 and p1 in Figure~\ref{themis}, was already visible on 9 November at around 16:00 UT; its PIL was oriented in the E--W direction. 
Later on, part of polarity p1, called p1$^\prime$, moved towards the West (see the movie {\sf magnetic-evolution.mpg} and Figures~\ref{EMF} and ~\ref{themis}(a)). A second bipole [n2--p2] was observable on 10 November at 00:00 UT, while a third bipole  [n3--p3] appeared on the same day at around 09:30 UT. The last bipole [n4--p4] increased in strength between 10 and 11 November. 
Several positive and negative small patches were clearly seen between n1 and p1 and between n4 and p4 during the first emergence stages forming a ``sea-serpent'' pattern \citep[\eg ][]{Pariat04}.
The first bipole [n1--p1] was oriented in a NE to SW direction, while [n2--p2] and [n3--p3] were oriented E--W, and finally [n4--p4] was mainly oriented S--N. This change in the successive bipole orientations created a complex and untypical active region.  As time advanced, there was a tendency of the polarities to cluster according to their sign; in particular, forming the following polarity (n1, n2, n3 and n4). 
 
\begin{figure} 
\vspace*{-9.5cm}
\centerline{\includegraphics[width=\textwidth,clip=]{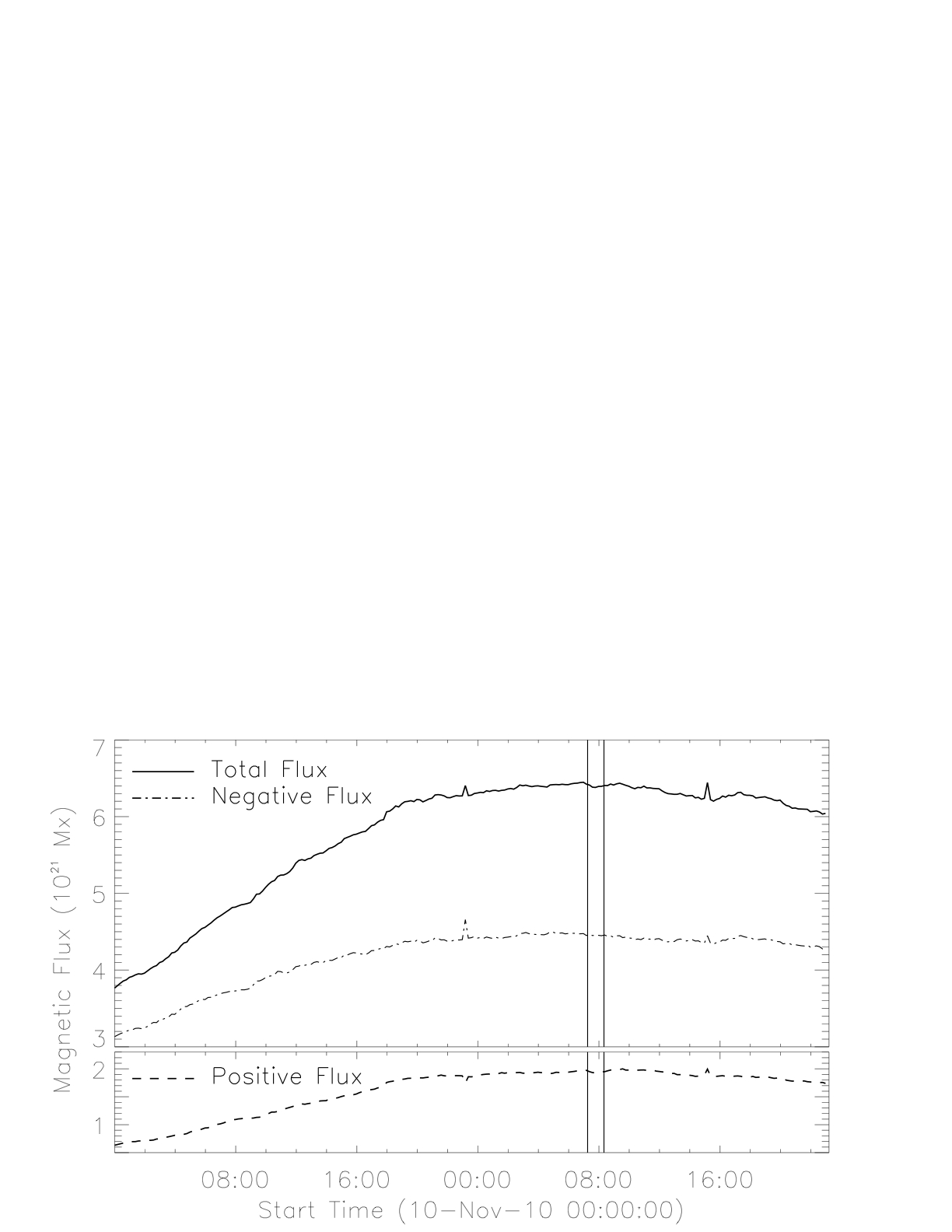}}
\caption{Magnetic-flux evolution in AR 11123. The flux is computed for the same field of view shown in Figure~\ref{themis}(a) within the green rectangle. The existence of a negative background is evident in the figure. The two vertical lines indicate the start and end times of the series of events discussed in detail in Section~\ref{S_Fil_Eruption}.} 
\label{flux-evol}
\end{figure}

Both HMI and THEMIS provide the vector magnetic field of this region on 11 November. We have compared the vector field obtained with these two instruments and found similar results concerning the field shear, although the strength of the horizontal component is different (Figures~\ref{themis}). The transverse components of magnetic fields suffer from a so-called 180$^\circ$ ambiguity. The ambiguity has been resolved using an improved version of the minimum-energy method \citep{Metcalf94,Metcalf06,Leka09}. When an AR is not located at the central meridian, projection effects distort the geometric shape and change the observed field components (line-of-sight and transverse) from the heliographic field components (vertical and horizontal). We have removed the projection effects from the field measurements in AR 11123, located at a heliocentric angle of about 24$^\circ$, using the formulae of \citet{Gary90}. 
The presence of a negative environment, in which AR 11123 was born, is clear in Figure~\ref{flux-evol}, where we show the variation of the positive and negative flux. It is also evident that the total unsigned flux increases by around 70~\% in less than one day. 

\begin{figure} 
\centerline{\includegraphics[width=\textwidth,clip=]{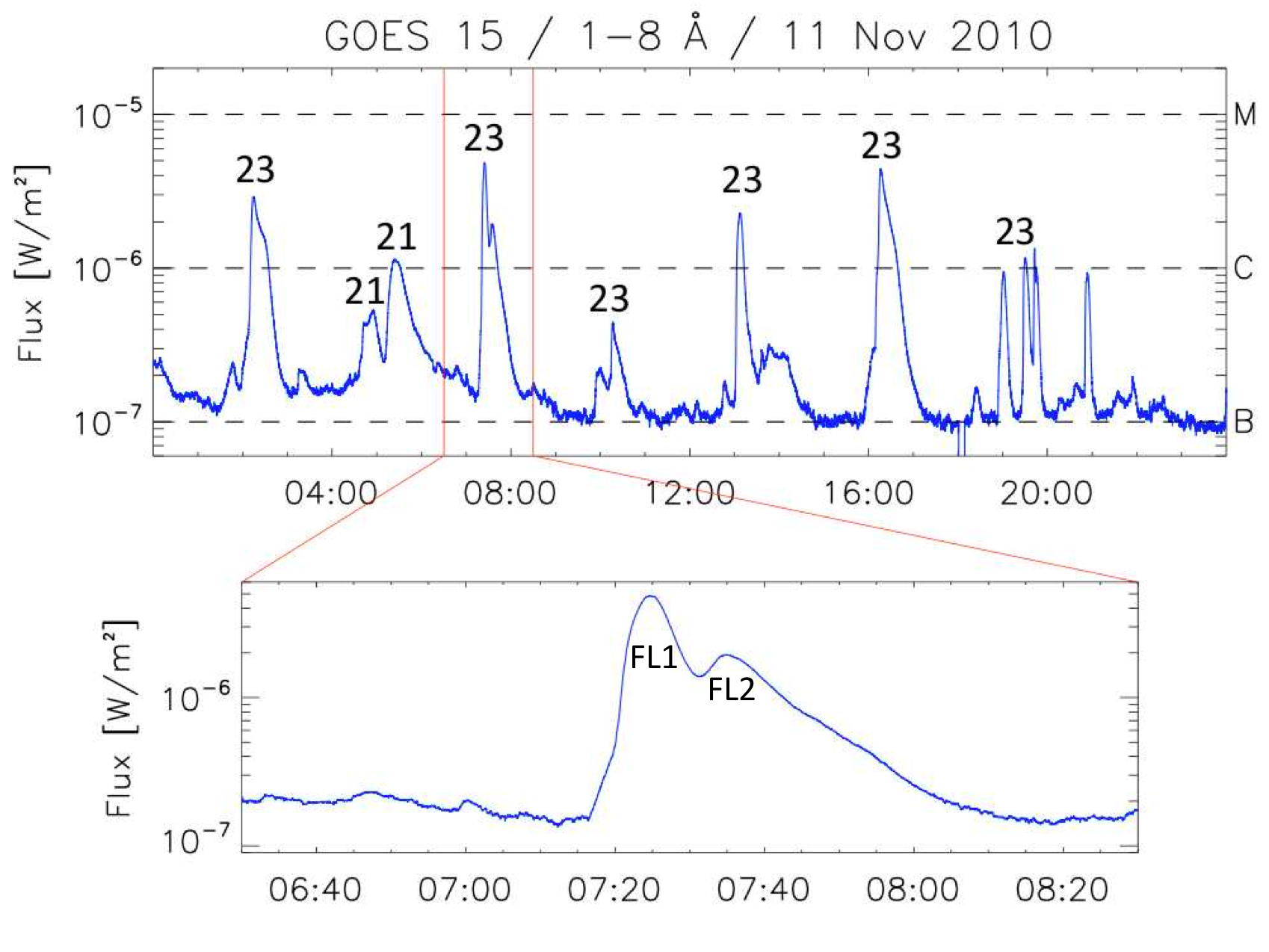}}
\caption{Soft X-ray flux as recorded by GOES in the 1 -- 8~\AA~energy band. The top curve shows the flares on 11 November 2010, while the bottom curve is an enlargement at the time of the events described in Section~\ref{S_Fil_Eruption}. The numbers above the different flare maxima in the top curve correspond to the AR where they occur (21 stands for AR 11121 and
23 for AR 11123). The labels in the bottom enlargement indicate the events that we discuss in detail in Section~\ref{S_Fil_Eruption}.} 
\label{GOES}
\end{figure}

\subsection{The Activity in the Complex}
\label{S_Activity}
 
Several X-ray flares of classes B and C took place in 
ARs 11121 and 11123.  In particular, five C class flares occurred in AR 11123 on 10 November and five C and one B class 
flares on 11 November (Figure~\ref{GOES}); the series of small flares around 19:23 UT took place in a northern active region. The location of the flares was recurrent in the new AR 11123 and some of them were connected to AR 11121.
The flare with maximum at 07:25 UT in the GOES soft X-ray curve (Figure~\ref{GOES}) was associated with a small CME with low speed, around 300 km s$^{-1}$ \cite{Schmieder13a}.   

Many absorbing dark structures were observed in AR 11123. These were not visible in the H$\alpha$ survey obtained with the spectroheliograph of the Paris Observatory, Meudon, because they were small and embedded in plages.  In AIA 304 \AA\ images, the filaments appeared either in absorption or in emission when they erupted (Figure~\figAIA ).  
Three filaments were detectable as shown in Figure~\figAIA (a): F1 along the PIL between n2--n4 and p1--p1$^\prime$, F2 along the PIL between n3 and p4, and F3 between p2 and n0W (the western negative zone of AR 11121). All of them were active-region filaments with lengths between 20 and 30 Mm, lying close to the 
chromosphere.
 
Let us describe briefly each flare in AR 11123.
The events occurring in the AR complex, up to 12:40 UT on 11 November, are shown in a movie with low temporal cadence (four minutes) that combines three AIA filters ({\sf 11Nov2010-AIA-171-211-304.mpg}) and is attached as online material.
The C2.9 flare, with maximum at 02:14 UT in Figure~\ref{GOES}, was associated with a small eruption starting at 02:00 UT between n3 and p1.  
We observe a semicircular brightening to the SE of the region between 02:06 and 02:12 UT, located partially on n0E and extending to n0S (see a snapshot in Figure~\ref{qsls-extra}(b)).
Bright post-flare loops were seen to develop at around 02:20 UT and later. There was no apparent change in the filaments during this event. At around 07:16 UT a filament eruption, accompanied by a two-ribbon flare, started and was followed by a confined flare; these events and their context will be discussed in Section~\ref{S_Fil_Eruption}.
The flare, whose GOES curve peaked at 10:16 UT, started with the brightening of loops close to n4 followed by a semicircular brightening towards the South, similar to the one observed during the 02:14 UT flare.
At 13:01 UT, filament F1 started moving slowly towards the South and we see two ribbons at both sides of the PIL between n2 and p1 with no clear eruption. They were visible until 13:30 UT. Between 15:53 and 16:27 UT, a new flare of class C4.3 occurred in AR 11123. This event was observed in H$\alpha$ as a two-ribbon flare by the {\it H$\alpha$ Solar Telescope for Argentina} \citep[HASTA: ][; see Figure~\ref{qsls-extra}(c)]{Borda02}, located at both sides of the PIL where filament F2 lay.
Finally, after some flares occurring in the northern hemisphere, there was a flare in AR 11123 with maximum at 19:30 UT. It was again a two-ribbon flare partially involving  F3 with no evident eruption.
 
\subsection{The Context and the Sequence of Events at $\approx$ 07:16 UT in AR 11123}
\label{S_Fil_Eruption}

A series of movies in the same AIA channels as those in the combined one, but separately and with the highest AIA temporal cadence, are included as online material. These cover the period from 06:00 UT to 08:00 UT and are called {\sf 11Nov2010-AIA-171.mpg}, {\sf 11Nov2010-AIA-211.mpg} and {\sf 11Nov2010-AIA-304.mpg}. Snapshots of the emission evolution in 304 \AA\ are shown in Figure~\figAIA . 

Filament F1 was seen as a dark feature (see Figure~\figAIA (a)) up to $\approx$ 06:58 UT in movie {\sf 11Nov2010-AIA-304.mpg}. Its brightness increased during almost two minutes, without any evident eruption, and appeared dark again by $\approx$ 07:04 UT. Around that time, an elongated thin brightening was seen along n0S extending in a SE to NW direction. Its brightness increased in 304 \AA\ and was maximum at $\approx$ 07:17 UT. A snapshot of this evolution is shown in Figure~\figAIA (b), where other smaller brightenings are seen on n2 and the eastern portion of n4 (see also 
Figure~\ref{qsls-N1-N3}(b)). From $\approx$ 07:08 to $\approx$ 07:17 UT, filament F1 appeared as a dark feature in expansion. A set of short loops, partially masked by filament F1, was seen in 304, 171 and 221 \AA\ connecting n0S to p1$^\prime$ (Figure~\figAIA (b)). Based on our topological analysis (see Sections~\ref{S_Model_Topo_Null} and~\ref{S_Model_Topo_QSL}), we infer that the brightenings observed on n0S and n2--n4 could be precursors to the two-ribbon flare that developed later along the PIL between n2 and p1.

The GOES soft X-ray curve (Figure~\ref{GOES}) started rising at 07:16:16 UT and its slope became steeper around 07:20 UT. Along this period filament F1 became bright and erupted. Two flare ribbons were visible in 304 \AA\ around 07:19 UT at both sides of the PIL where filament F1 lay. At later times, the emission in that AIA band saturated and the ribbons were no longer discernible (see Figure~\figAIA (c)). The GOES curve reached a maximum that corresponds to class C4.7 at 07:25 UT. We refer to this first event as FL1. The ejected material followed a curved path towards the SW (see Figure~\figAIA (d)) as tracing the shape of very large-scale loops connecting to the westernmost edge of the preceding polarity in AR 11121. These large-scale loops are better seen in the movie combining the three AIA channels at times without activity.

A second maximum appears in GOES curve at 07:35 UT, we name this event as FL2. 
As filament F1's eruption continued, a roundish ribbon (on n1, n3, n2, and n4 to the East and n0w to the West) with a bright central bar (on p3 and p2) started to be seen (see Figures~\figAIA (d) and (e)); these indicate the location of event FL2. These regions appeared clearly in emission at around 07:32 UT and later on, when the brightness due to the eruptive flare FL1 started fading. Loops were seen connecting these brightenings (see Figure~\figAIA (e)). FL2 is a new confined flaring episode probably induced by the filament eruption. 
In later images ($\approx$ 07:35 UT) a ``post-flare'' arcade was clearly seen along the PIL, where filament F1 previously lay and flare FL1 was initiated. By 08:00 UT the global emission was decreasing and, finally, it faded.

\begin{landscape}
\begin{SCfigure} 
\centering
\hspace*{-1.4cm}
\includegraphics[width=1.25\textwidth,clip=]{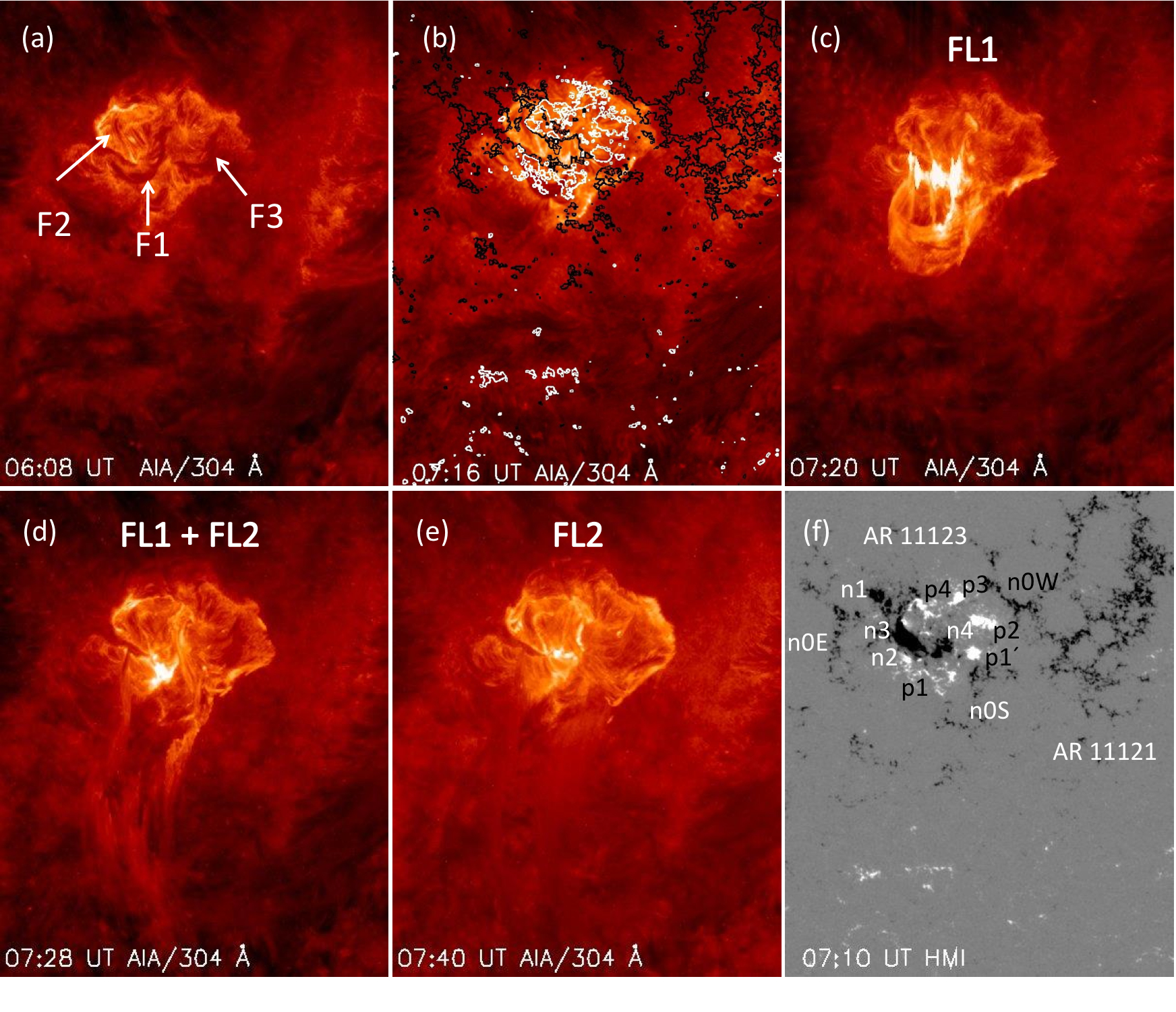}
\vspace*{-0.8cm}
\caption{
   (a) Filaments F1, F2 and F3, embedded in AR 11123, observed in absorption on 11 November 2010 at 06:08 UT. 
   (b)--(e) Different stages of the eruption of filament F1, low-energy event preceding (or precursor to) FL1 and flare (FL1 and FL2) brightenings. The upper chromosphere images are observed with AIA at 304 \AA. Isocontours ($\pm$ 50, $\pm$ 100 G, white (black) continuous lines for positive (negative) field values) of the HMI image shown in panel (f) have been added as a guide in panel (b). 
   (f) HMI line-of-sight magnetic field covering the same field of view as the AIA images.
The polarities involved in the studied events are labelled. For clarity, these labels are placed at the side of the polarities in AR 11123. A detailed description of the evolution illustrated in these images is provided in Section~\ref{S_Fil_Eruption}. The size of each panel is $240''$ ($300''$) in the horizontal (vertical) direction. The centre of each panel in heliographic coordinates is located at [-165$''$,-485$''$]; Sun's centre is at [0$''$,0$''$].}
\label{eruption}
\end{SCfigure}
\end{landscape}

\begin{figure} 
\vspace*{-2.5cm}
\centerline{\includegraphics[width=\textwidth]{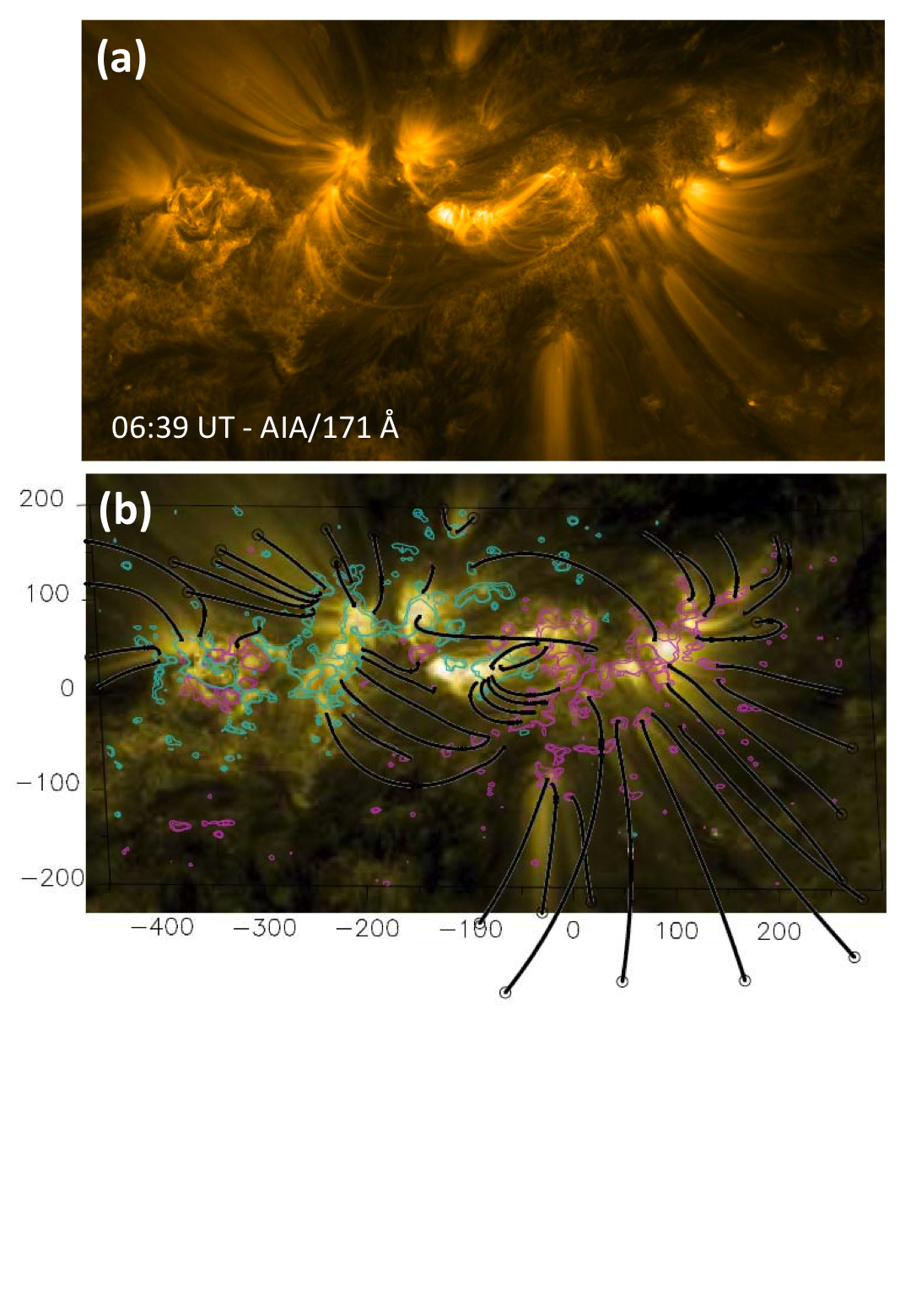}}
\vspace*{-4.3cm}
\caption{Coronal magnetic-field model of the AR complex.  
(a) AIA 171 \AA\ image. The size of this panel is $690''$ ($390''$) in the horizontal (vertical) direction. Its centre in heliographic coordinates is located at [-50$''$,-440$''$]; Sun's centre is at [0$''$,0$''$].
(b) Field lines (black continuous lines), computed from the LFFF model,
that best fit AIA loops for a value of $\alpha$ = 3.1 $\times$ 10$^{-3}$ Mm$^{-1}$.  The field lines ending in an open circle reach the 3D box selected for the figure (only its base is shown).  The axes in this panel 
are in Mm and the isocontours of the field correspond to $\pm$ 50, $\pm$ 100 G in
continuous magenta (blue) style for the positive (negative) values. These are overlaid on the AIA image shown in (a).}
\label{linear1}
\end{figure}

\section{Magnetic-Field Model and Null Points}
\label{S_Model_Topo_Null}

\subsection{Large-scale LFFF model}
\label{S_Large_Scale}

To understand the origin of the events described in the previous section and their relation to the 3D magnetic structure of the AR complex, we have extrapolated the photospheric line-of-sight field, observed with HMI, to the corona. We first compute the coronal magnetic field under the LFFF (or constant-$\alpha$) assumption following the work of \citet{Demoulin97}, which is based on a fast Fourier transform method proposed by \citet{Alissandrakis81}. Although this model cannot take into account the photospheric-current distribution, it has proven to be fast and efficient to compute
the magnetic-field topology and compare it with observed active events (see references in Section~\ref{S_Introduction}).
The model takes into account the transformation of coordinates from the location of the region on the Sun, as seen by an observer at 1 AU, to the local solar coordinates.

To determine the free parameter [$\alpha$], on which the LFFF model depends, we have proceeded as did \citet{Green02}. A discussion of the limits of the extrapolation method is presented by \citet{Demoulin97}. We have taken an AIA image at 06:39 UT in the 171 \AA\ channel, when no flare or ejection was observed, to compare the computed magnetic-field lines to AIA coronal loops (see Figure~\ref{linear1}). The HMI magnetogram taken as boundary condition for the model is the closest in time to this AIA image. 
Figure~\ref{linear1}(b) includes the computed field lines on the HMI line-of-sight magnetogram in the same field of view as the AIA image. These field lines correspond to a model with $\alpha$ = 3.1 $\times$ 10$^{-3}$ Mm$^{-1}$. A positive value for $\alpha$ is consistent with the tongues observed during the emergence of one of the bipoles in AR 11123 (see Section~\ref{S_Evolution}) and with the photospheric magnetic-flux distribution of the two previous emergences in AR 11121 (Figure~\ref{HMI_long-term}).

\begin{landscape}
\begin{figure} 
\centering
\includegraphics[width=1.45\textwidth]{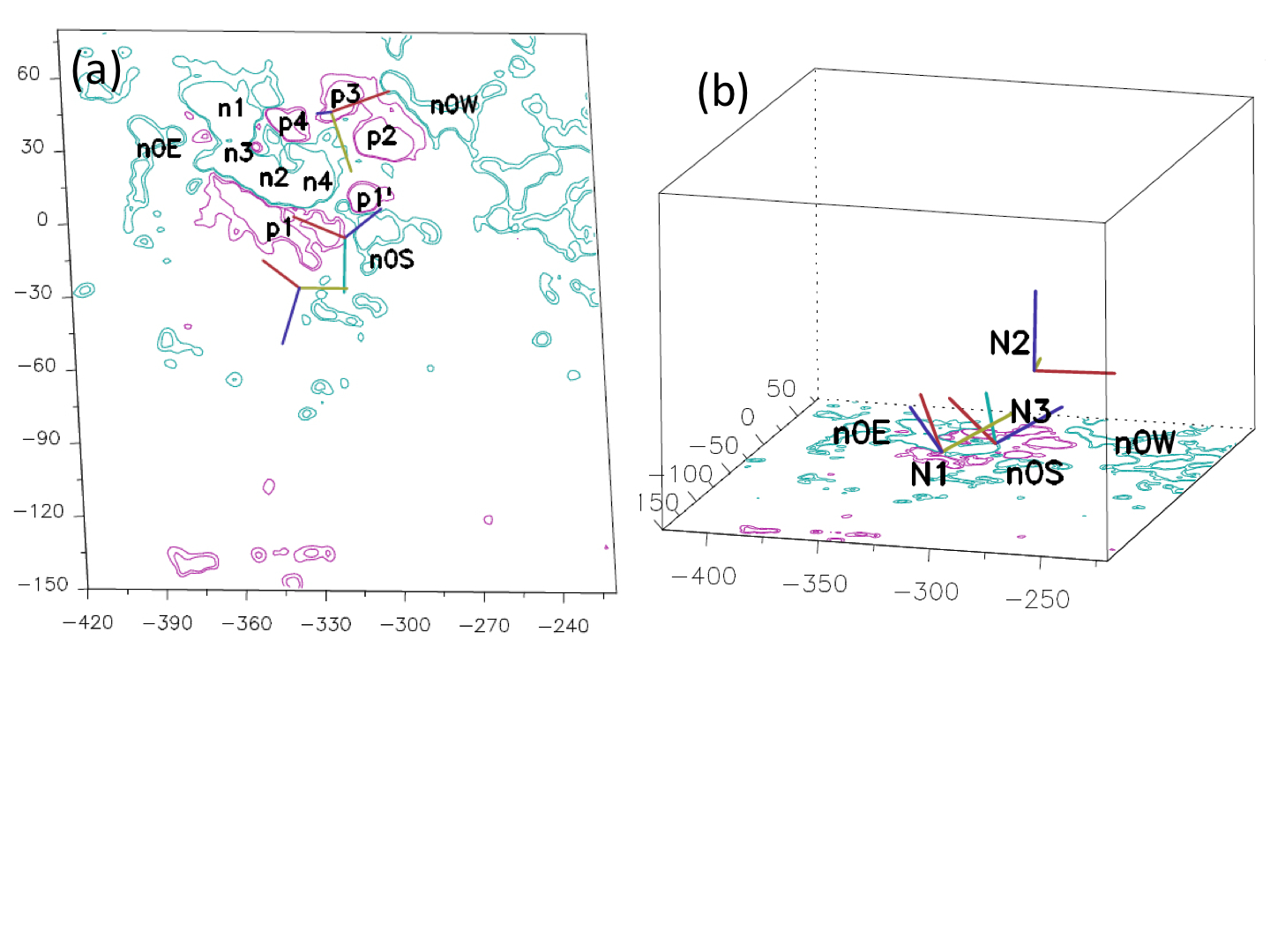}
\vspace*{-4.2cm}
\caption{Location of the three null points associated with the magnetic configuration of AR 11123 on 11 November 2010. 
(a) Observer's point of view showing the three null points. 
For reference, we have numbered the polarities as in Figure~\ref{themis}.  
(b) Perspective view selected to illustrate the height of the nulls above the photosphere, which are numbered N1, N2 and N3. Some of the polarities are labelled as in Figure~\ref{themis}. In both panels, all axes are in Mm and the isocontours of the field correspond to $\pm$ 50, $\pm$ 100 G in
continuous magenta (blue) style for the positive (negative) values. The three segments at the locations of the nulls
correspond to the direction of the three eigenvectors of the Jacobian matrix. The colours of these segments indicate the magnitude of the corresponding eigenvalue, red (yellow) corresponds to the largest (lowest) positive eigenvalue in the fan plane and blue to the spine eigenvalue for a positive null point (N1 and N2 in our study). In the case of a negative null (N3 in our study),
light blue (dark blue) corresponds to the largest (lowest) negative eigenvalue in the fan plane and red to the spine eigenvalue. From now on, the field of view of each figure will be chosen according to the magnetic structures that we want to show. As a guide, we will label in each figure some of the polarities as in Figure~\ref{themis}.}
\label{threenulls}
\end{figure}
\end{landscape}

\begin{landscape}
\begin{figure} 
\centering
\includegraphics[width=1.6\textwidth]{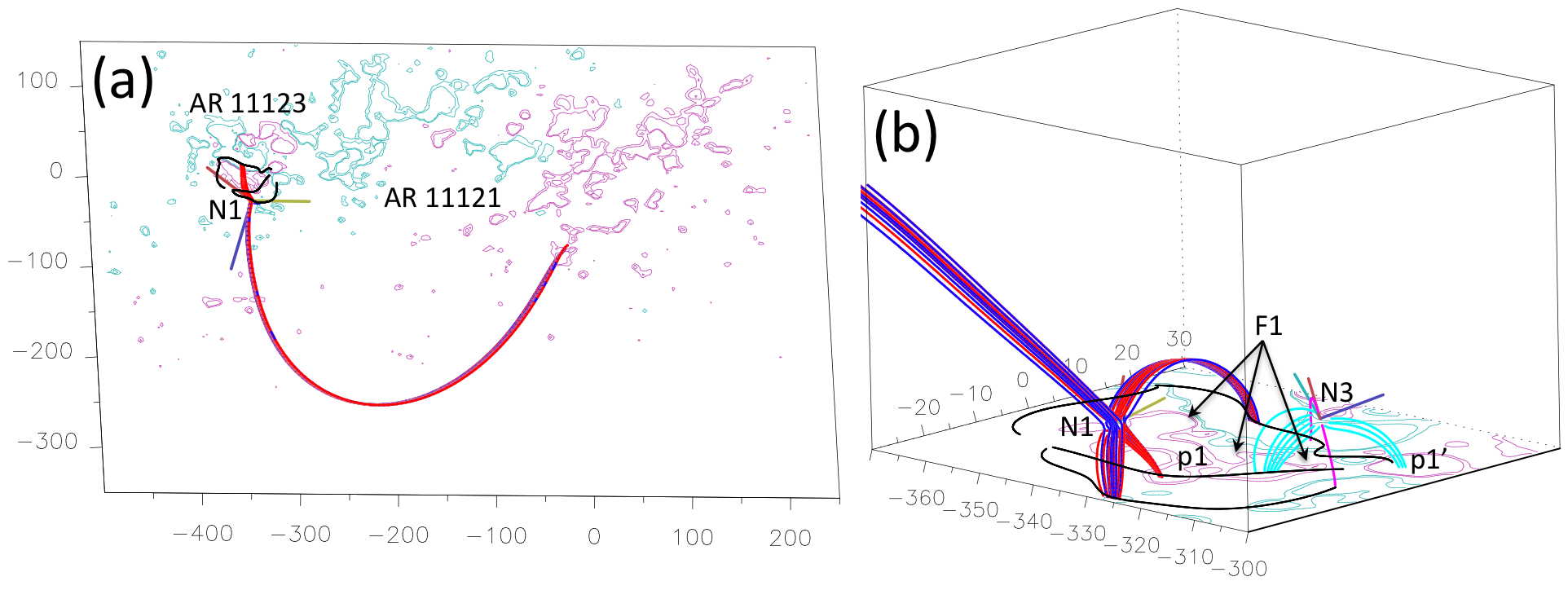}
\caption{
Coronal magnetic-field model in the close vicinity of magnetic null points N1 and N3. 
(a) Large-scale connection of the spine of null N1. The set of N1 fan field lines, following the upper spine, connected to AR 11121 are shown in the observer's point of view.      
(b) Close-up view showing sets of field lines drawn from the close neighbourhood of null points N1 and N3. In the case of N1, field lines represent pre-reconnected (in blue colour) and reconnected ones (in red colour), as inferred from the observed evolution described in Section~\ref{S_Fil_Eruption} and the interpretation in Section~\ref{S_QSLs_Fil-erup}. For N3, the spine is drawn in pink and a set of fan field lines in light blue. The arrows in panel (b) point to the location of filament F1 along the PIL between n2-n4 and p1-p1$^{\prime}$.
In both panels, the black continuous lines correspond to the photospheric trace of QSLs (see Section~\ref{S_Model_Topo_QSL}). 
All axes are in Mm and the isocontours of the field correspond to $\pm$ 50, $\pm$ 100 G in
continuous magenta (blue) style for the positive (negative) values.}
\label{null1-3}
\end{figure}
\end{landscape}

\subsection{Magnetic Null points in AR 11123 in the LFFF Model}
\label{S_Null_Points}

We find three magnetic null points associated mainly to the magnetic configuration of AR 11123 (Figure~\ref{threenulls}). These null points are located 
at a height above the photosphere of around 8.0 Mm (N1 and N3) and of around 25 Mm (N2). 

The vicinity of a null point can be described by the linear term in the local Taylor expansion of the magnetic field. Diagonalizing the Jacobian field matrix gives three eigenvectors \cite{Molodenskii77}.  The divergence-free condition on the field imposes that the sum of the three eigenvalues vanishes ($\lambda _1 +\lambda _2 +\lambda _3 =0$). Furthermore, for cases in which the  magnetic field is in equilibrium with the plasma [$\vec{j} \times \vec{B} = \vec{\nabla} P$], the eigenvalues are real \cite{Lau90}. That is to say,  
two eigenvalues have the same sign, which is opposite to that of the third eigenvalue. The presence of a null point divides
the coronal volume into two connectivity domains, separated by the fan surface. In each domain a spine is present, in what follows we
call the lower (upper) spine to the one below (arising from) the fan surface.
These two spines are defined by the two field lines that start at an infinitesimal distance from the null in directions parallel and anti-parallel to the eigenvector associated with the eigenvalue having a different sign. The fan surface is defined by all of the field lines that start at an infinitesimal distance from the null in the plane defined by the two eigenvectors associated with the eigenvalues having the same sign \citep[see ][, for more details]{Greene88,Lau93,Longcope05b}.

The two lower magnetic nulls (N1 and N3) have an opposite magnetic structure; N1 has two positive eigenvalues and
one negative \citep[positive null, \eg\ ][]{Longcope05b} and {\it vice versa} for N3
(negative null). The ratio of the eigenvalues in the fan plane is 7.2 for N1 and 2.8 for N3, 
respectively. Figure~\ref{null1-3}(b) shows a set of field lines traced from the close vicinity of both nulls.  
As we will discuss in Section~\ref{S_QSLs_Fil-erup}, reconnection at N1 is at the origin of the low-energy event preceding (or precursor to) flare 
FL1 and it may play a key role in the ejection of filament F1. Following our interpretation, we have used blue for field lines before reconnection and red for field lines after reconnection at N1. The upper spine at N1 connects to a positive-field region in AR 11121 (Figure~\ref{null1-3}(a)).  
The role of N3, whose magnetic structure is more localised, is not clear during the events described in Section~\ref{S_Fil_Eruption} because the emission associated with reconnection at this null is masked by that of flares FL1 and FL2 and the eruption of filament F1. That is why we are not using colours to distinguish field lines before or after reconnection (Figure~\ref{null1-3}(b)) associated with this null.  It is also clear, from Figure~\ref{null1-3}(b), that filament F1 lay below a set of N1 and N3 fan field lines that extend above the PIL between n2--n4 and p1--p1$^{\prime}$. This is confirmed by the NLFFF model that we analyse in Section~\ref{S_Small_Scale}.  

\begin{landscape}
\begin{figure} 
\centering
\includegraphics[width=1.5\textwidth]{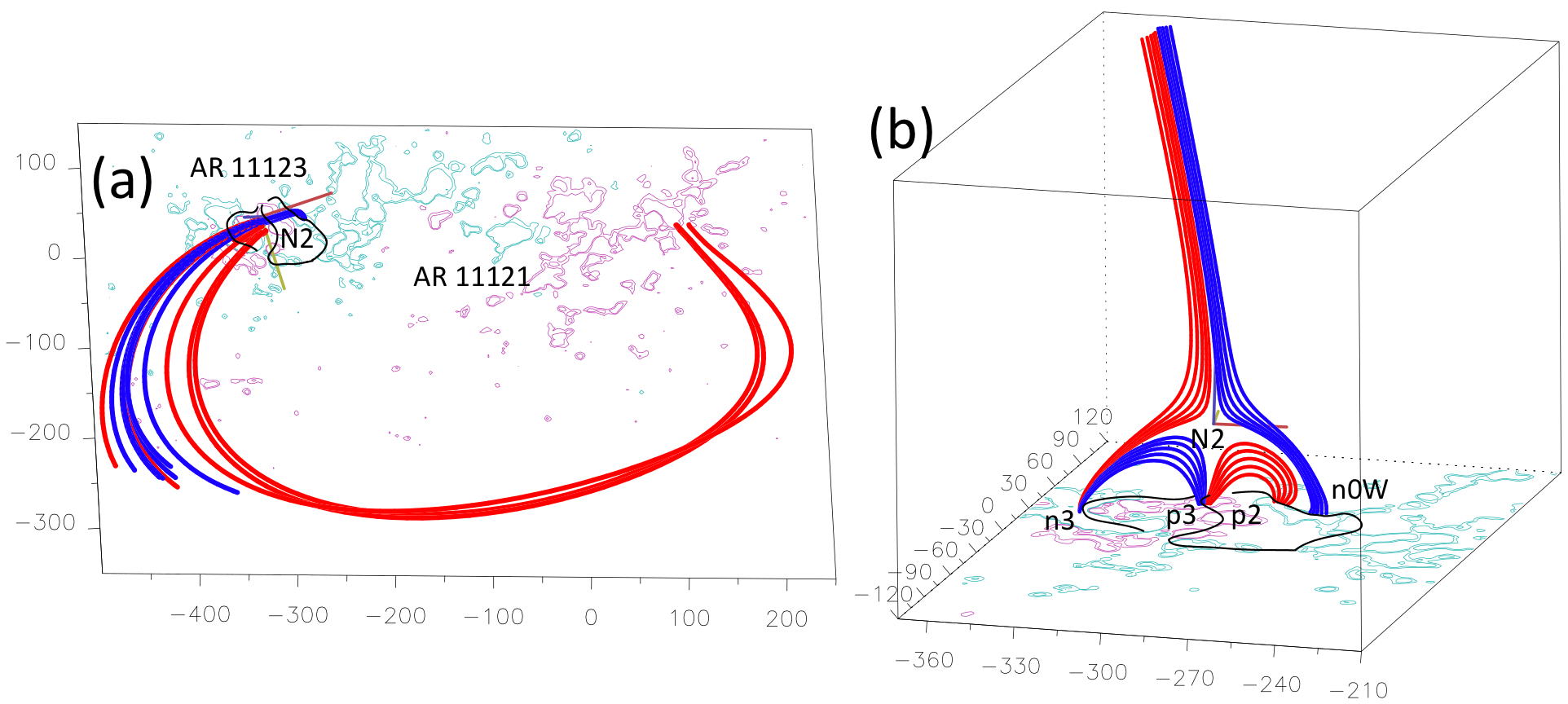}
\caption{Coronal magnetic-field model in the close vicinity of magnetic null point N2.
(a) The set of N2 fan field lines following the upper spine connected to AR 11121 are shown in the observer's point of view. 
(b) Close-up view showing two sets of field lines representing the pre-reconnected lines (in blue colour) and reconnected lines (in red colour), as inferred from the observed evolution described in Section~\ref{S_Fil_Eruption} and our interpretation in Section~\ref{S_QSLs_FL2}. 
In both panels the black continuous lines correspond to the photospheric trace of QSLs (see Section~\ref{S_Model_Topo_QSL}). All axes are in Mm and the isocontours of the field correspond to $\pm$ 50, $\pm$ 100 G in
continuous magenta (blue) style for the positive (negative) values.}
\label{null2}
\end{figure}
\end{landscape}

\begin{landscape}
\begin{figure} 
\centering
\includegraphics[width=1.65\textwidth]{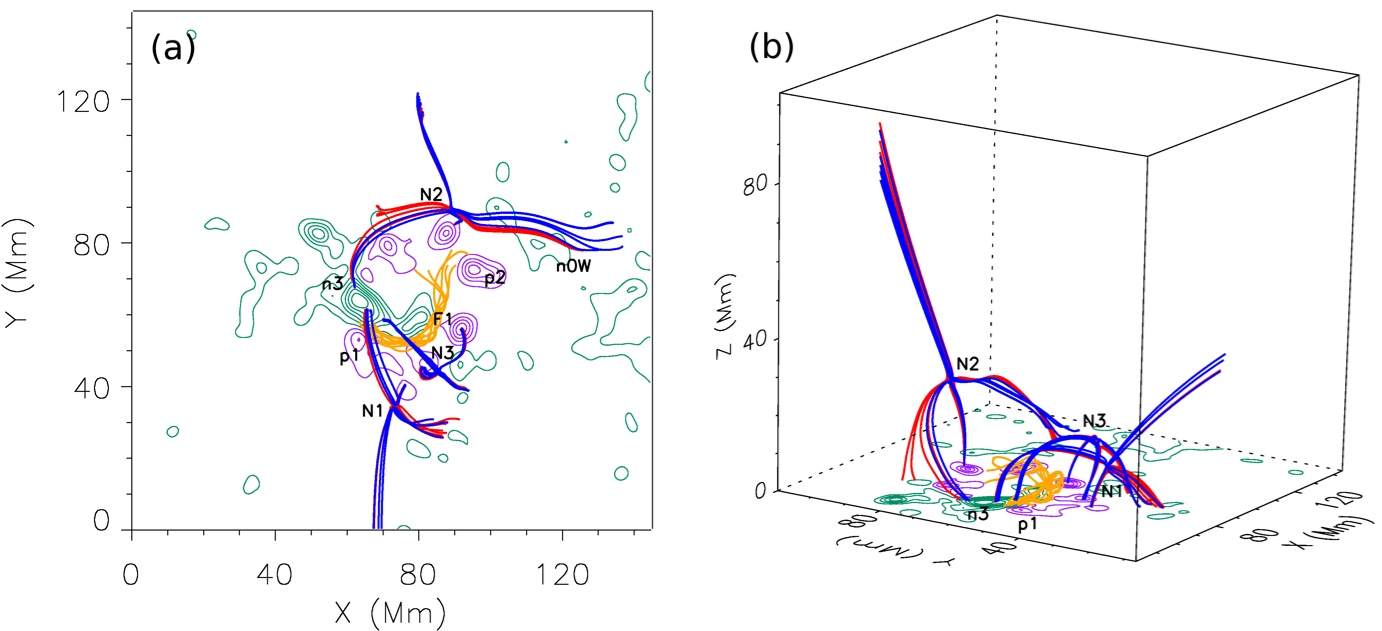}
\caption{Magnetic-field model of AR 11123 in the vicinity of the null points in the NLFFF approximation.
(a) Point of view in which the region is located at central meridian passage.
Blue and red field lines represent the magnetic domains before and after the magnetic reconnection, respectively, following the convention in Figures~\ref{null1-3} and~\ref{null2}. Orange field lines depict the magnetic-field structure of filament F1 which lies along the PIL between n2--n4 and p1--p1$^{\prime}$ (see Figures~\figAIA  and~\ref{null1-3}(b)). 
(b) Point of view selected to better show the 3D structure of the field lines.
In both panels the magnetic-field contours correspond to the photospheric vertical field component in the range [-1600,1600]~G, which is divided into 12 levels with a difference of 291 G between two adjacent levels.}
\label{nlfff}
\end{figure}
\end{landscape}

Magnetic reconnection at null N2 plays a crucial role during flare FL2 (see Section~\ref{S_QSLs_FL2}). For this particular null, one of the fan eigenvalues is much larger than the other ($\approx$ 14 times larger). Then, if we start computing field lines very close to the null, as for N1 and N3, most field lines in the fan plane bend in the directions parallel or anti-parallel to the eigenvector with the largest eigenvalue, instead of tracing the full fan. Consequently, all computed field lines stay nearly in one plane  (Figure~\ref{null2}(b)). As in the case of N1,  
and consistently with the evolution of flare FL2, we have drawn in blue (red) field lines before (after) magnetic reconnection at the null N2. 
Field lines from this set, drawn along the upper spine, connect away from AR 11123 as for N1 (Figure~\ref{null2}(a)).

\subsection{Small-scale NLFFF Model and Null-point Locations}
\label{S_Small_Scale}

We also model the small-scale structure of AR 11123 in the NLFFF approach. The vector magnetic field, shown in Figures~\ref{themis}(a) and (b), is adopted as the lower boundary condition for the model after correcting for projection effects. 

The NLFFF approach requires that vector magnetic fields on a closed volume have to satisfy the force-free and torque-free conditions. Since the magnetic-field vector is only available on the photospheric boundary, the force-free and torque-free conditions are required to be satisfied in a well-isolated region. We
adopt the method of \citet{Wiegelmann06} to remove the net magnetic force and torque, and we apply it to the magnetogram that includes both ARs, 11121 and 11123. The flux-balance parameter (the ratio between the total magnetic flux and its total absolute value) is about 1~\%. The preprocessing method is used in this flux-balanced region. The optimisation method is applied to this pre-processed field to compute the NLFFF model \citep{Wheatland00,Wiegelmann04}. First, an NLFFF extrapolation with lower spatial resolution (a grid with a $4\times 4$ binning) is computed in the large field of view including ARs 11121 and 11123. Then the derived model is cut and interpolated
to serve as the boundary and initial condition for another NLFFF extrapolation with higher spatial resolution (a grid with a $2\times 2$ binning), but in the smaller field of view shown in Figure~\ref{themis}(a).

As in the LFFF model, we also find three null points in the NLFFF approach. The field lines in the vicinity of the three nulls are plotted in Figure~\ref{nlfff}. The heights of N1, N2 and N3 are 12, 22 and 14 Mm, respectively. The three nulls have the same sign as those of the LFFF model.
The ratio between the two eigenvalues for the fan eigenvectors are 8.6, 1.9 and 2.6 for N1, N2 and N3, respectively. We conclude that the heights and magnetic structure of the nulls found in the LFFF and NLFFF models are similar, which is a further confirmation that the three nulls are indeed present in the magnetic configuration of AR 11123. The main difference in the characteristics of the three null points is that for N2 the relation between the two eigenvalues for the fan eigenvectors is much higher in the LFFF than in the NLFFF approach; {\it i.e.} null N2 is an asymmetric (close to 2D) null point in the LFFF approach and it is more symmetric in the NLFFF model (compare Figure~\ref{null2} to Figure~\ref{nlfff}). For nulls N1 and N3, the relation between the two fan eigenvalues is closer in both modelling approaches. Problems in the NLFFF model, such as the fact that the divergence-free condition is not well satisfied, could be at the origin of these differences.   

Another key structure in the events that we analyse in detail can be found by the NLFFF model. This is the low-lying highly sheared set of magnetic-field lines (orange lines in Figure~\ref{nlfff}) that represents the magnetic structure of filament F1 (see Figure~\figAIA ). The highest part of this set of orange field lines reaches approximately 4.4 Mm and still lies under the fan field lines of N1 and N3.

\subsection{Relevance of the Null points in the Observed Flares}
\label{S_Implications_null}

Since the three magnetic null points have, in general, different eigenvalues in their fan planes, field lines passing close to any of them have a strong tendency to cluster around the direction of the eigenvector with the largest eigenvalue (Figures~\ref{null1-3}--\ref{nlfff}).  This contrasts with the shape and extension of the emission of the low-energy precursor to FL1 and that of flare FL2, described in Section~\ref{S_Fil_Eruption}. The brightenings from these events are not restricted to the portion of the separatrix that we can trace when only computing field lines from the close neighbourhood of the nulls.  A possibility to explore would have been to search for the presence of separators linking the nulls of opposite sign: N1 to N3 and/or N2 to N3. However, based on our past results that relate the magnetic-field topology to flares (see Section~\ref{S_Introduction}), we rather choose to explore a broader concept by computing QSLs. If a separator is present, it is a particular case of a hyperbolic flux tube (HFT) that characterises the core of QSLs \cite{Titov02}.

More precisely, we describe the relevance of null points N1 and N3 together with that of QSLs in Sections~\ref{S_QSLs_Fil-erup} and~\ref{S_QSLs_FL2} for both flares. Indeed, we recall that a separatrix of a null point is a special part of a QSL where the field-line connectivity is discontinuous.  Indeed, a separatrix is embedded in a QSL and a drastic change of connectivity occurs at both of its sides \citep{Masson09}. However, QSLs are a more general concept than separatrices, which are present in 3D magnetic configurations even without magnetic null points. 3D reconnection occurs at QSLs with the continuous slippage of magnetic-field lines, as deeply analysed in MHD simulations \citep[][, and references therein]{Janvier13}.  When a separatrix is present within a QSL, the slippage velocity simply becomes infinite when crossing the separatrix.  Then, in 3D magnetic configurations, reconnection at null points can be viewed as a particular case of reconnection within QSLs.  In order to shorten the description of the relationship between the coronal-field model derived in Section~\ref{S_Large_Scale} and the flare observations we describe together the implications of nulls and QSLs in Sections~\ref{S_QSLs_Fil-erup} and~\ref{S_QSLs_FL2}, emphasizing the role of the null points when appropriate.

\section{Quasi-separatrix Layers in the AR Complex}
\label{S_Model_Topo_QSL}

\subsection{Computing QSLs}
\label{S_QSLs_General}
   
The method to find QSLs in a magnetic configuration was first described by \citet{Demoulin96a}. QSLs were then defined using the norm, $N$, of the Jacobian matrix of the field-line mapping.
However, $N$ depends on the direction selected to compute the field-line mapping; therefore, $N$ has, in general, different values at both photospheric footpoints of a field line.  \citet{Titov02} solved this problem by showing that the squashing degree ($Q$) is independent of the mapping direction.  $Q$ is simply $N$ squared divided by the ratio of the vertical component of the photospheric field at both field-line footpoints. This ratio corrects for the natural expansion of a flux tube with variable field strength and magnetic-flux conservation. Then $Q$ includes only the distortion of the field-line mapping, independently of the field strength, and has the same values at both field-line footpoints. $Q$ can be also computed in the coronal volume and has the property of being constant along each field line.

QSLs are computed by integrating a huge number of field lines.  A key point is to use a very precise integration method since derivatives of the mapping
 are later computed to calculate $N$ and, {\it a posteriori}, $Q$. In order to decrease the computation time we use an adaptive mesh, as follows. First, $Q$ is computed on a coarse mesh, which has a spatial resolution comparable to that of the magnetogram.
Then this mesh is refined iteratively only around the locations of the largest values of $Q$.
The fraction of points retained at each iteration controls the computation speed and how much the finally 
calculated $Q$-map will extend towards the lower values of $Q$.  The iteration at a location is ended when the QSL is locally well resolved or, ultimately, when the limit of the integration precision is reached.  Such computations can be performed at the photospheric level (as in this article) and also within the full coronal volume \cite{Pariat12}.  As in previous work, we select a large value of $Q$ to show the extension of the QSL traces at the photospheric level.
 
In Sections~\ref{S_QSLs_Fil-erup} and~\ref{S_QSLs_FL2}, we show the trace of QSLs only for the LFFF model because the topology, calculated in Section~\ref{S_Model_Topo_Null}, is basically the same for the LFFF and the NLFFF extrapolations. Furthermore, the location of QSLs is strongly determined by the distribution of the field polarities at the photosphere in complex regions (see references in Section~\ref{S_Introduction}), which is the same for both models. However, as the LFFF model is not able to represent the highly sheared structure of filament F1, QSLs at both sides of the PIL where filament F1 lies, and that would trace the ribbons of flare FL1, will not be present.

\subsection{QSLs and the Precursor of Flare FL1}
\label{S_QSLs_Fil-erup}

Null points, their associated separatrices, and, more generally, QSLs are the expected locations where magnetic reconnection can occur efficiently in the corona to release the stored magnetic energy (Section~\ref{S_Introduction}).  The released energy is transported along field lines, by energetic particles and thermal conduction, toward the chromosphere, where it is deposited and we observed the flare ribbons.  Then, in this theoretical framework, we are expecting that the flare ribbons are located at the base of the QSLs.  Below, we compare such prediction, based on the magnetic extrapolation of the photospheric magnetic field, with the observed flare ribbons. Compared to previous studies, the challenge here is the complexity of AR 11123 magnetic configuration,
which was formed by the emergence of several non-parallel bipoles. This creates a complex coronal magnetic configuration, with three null points at a significant height and a complex ``web'' of QSLs. 

\begin{landscape} 
\begin{figure} 
\centering
\includegraphics[width=1.25\textwidth]{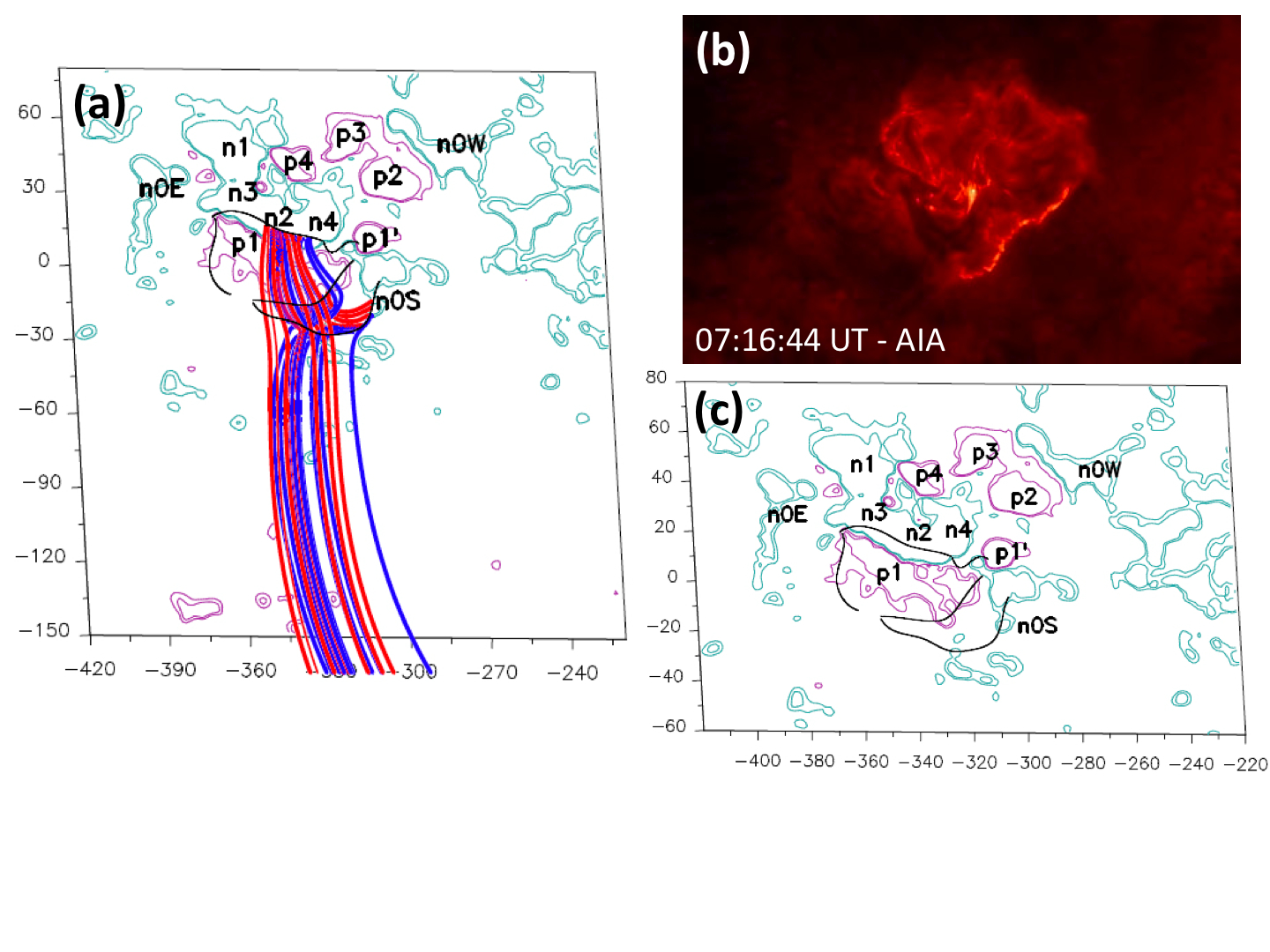}
\vspace*{-2.2cm}
\caption{Comparison of QSLs and brightenings of the low-energy precursor to flare FL1.
   (a) Photospheric trace of QSLs in AR 11123 computed on the polarities associated to null N1 and field lines originating in the QSLs. Blue and red field lines indicate pre-reconnected and reconnected loops. The field lines that leave the drawn box correspond to large-scale loops connecting to a positive polarity at the South East in AR 11121 (see Figure~\ref{null1-3}(a)).
   (b) An AIA 304 \AA\ image in which an elongated brightening along polarity n0S and other minor ones on polarities n2 and n4 are present. The size of this panel is 240$''$ (150$''$) in the horizontal (vertical) direction. Its centre is located at [-175$''$,-430$''$] in heliographic coordinates; Sun's centre is at [0$''$,0$''$].
   (c) Photospheric QSL traces and magnetogram.  This panel is a section of panel (a) and the field of view is the same as in panel (b). There is a close resemblance between the shape of the southernmost QSL and a section of the AIA brightening to the South of n0S, while the emission on n2 and n4 lies along the northernmost QSL. 
The shape of the western extension of the elongated band follows the shape of a section of the QSLs in Figure~\ref{qsls-N2}, see Section~\ref{S_QSLs_Fil-erup} for a plausible explanation.  
   In panels (a) and (c), all axes are in Mm and the isocontours of the field correspond to $\pm$ 50, $\pm$ 100 G in
continuous magenta (blue) style for the positive (negative) values.
}
\label{qsls-N1-N3}
\end{figure}
\end{landscape}

The photospheric trace of QSLs (black continuous lines) located along the polarities associated to null N1 is shown in Figure~\ref{qsls-N1-N3}(a), together with a set of field lines computed starting integration at both of their sides. 
The value of $Q$ is extremely high as these QSLs have separatrices within them; in particular, 
$Q$ $\ge$ 10$^{10}$ for the traces shown in Figures~\ref{qsls-N1-N3}(a)\,--\,(c). For the sake of clarity, Figure~\ref{qsls-N1-N3}(c) illustrates the QSLs on a section of the magnetogram in the left panel that covers the same field of view of the AIA image in Figure~\ref{qsls-N1-N3}(b). The shape of the QSLs can be easily compared to the shape of the 304 \AA\ emission. The southern portion of the elongated brightening or band clearly agrees with the shape of the southernmost QSL. Furthermore, a set of smaller AIA brightenings are located on polarities n2 and n4 along the QSL lying there. This indicates that magnetic reconnection at null N1 and its associated separatrices/QSLs could be at the origin of this set of brightenings preceding flare FL1.     

Taking into account the computed topology and the evolution of the events discussed in Section~\ref{S_Fil_Eruption}, the role of magnetic reconnection at null N1 can have two different interpretations in relation to the eruption of filament F1 and accompanying flare FL1. 

First, field lines connecting polarities n2--n4 to p1 can reconnect at the location of null N1 with the large-scale lines having footpoints in n0S and the positive polarity to the SE of AR 11121
(Figure~\ref{null1-3}(a)), which we will call pSE. Both sets of field lines are shown in blue in Figure~\ref{qsls-N1-N3}(a). As a result of this process, the reconnected field lines would link polarities n0S to p1 and n2--n4 to pSE. These have been drawn in red in Figure~\ref{qsls-N1-N3}(a). This reconnection process starting at null N1 neighbourhood and its separatrices, where currents are expected to be the strongest, would evolve to involve the corresponding QSLs \cite{Masson09}. In this case, we expect to see brightenings on n0S and p1 and n2--n4 and pSE, where the QSLs are located. Some of these brightenings appear in Figure~\ref{qsls-N1-N3}(b) (see also the movie {\sf 11Nov2010-AIA-304.mpg}); however, the dark filament F1 in expansion covers the emission on p1 and partially to the East of n0S. We should also observe an enhancement on pSE, but if an equal amount of energy is injected in both small-scale and large-scale reconnected loops, then this is very unlikely due to the dilution of the energy deposited in a large volume, as seems to be the case when looking at the movies attached as online materials; even more, if we consider that this is not as energetic event as flares FL1 and FL2. In this case, reconnection at N1 would decrease the tension of the field overlying filament F1 and facilitate its eruption as in a break-out model 
\cite{Antiochos98,Antiochos99}, but occurring in a lateral connectivity cell \cite{Aulanier00}. 
This low-energy event would be a precursor to FL1 in a similar way to field reconnection associated with the four kernels observed before the X17 flare on 28 October 2003 \cite{Mandrini06}. 

Second, it is also possible that the filament eruption is driven by a tether-cutting process \cite{Moore01}, as is suggested by the fact that the filament apparently moved upward and became bright at $\approx$ 06:59 UT for a few minutes. However, no eruption occurred and filament F1 was seen dark again from $\approx$ 07:04 UT to $\approx$ 07:17 UT, when the elongated band on the southernmost QSL evolved increasing its brightness. The filament started to expand again from around 07:08 UT and later erupted. In this case the direction of reconnection was the same as in the first case, but it was forced at null N1 by the filament itself as it expanded. From a conservative point of view, we feel that the observations are not conclusive in supporting one or the other interpretation.

Next, we point out that the westernmost sections of the QSLs located on n4 (mainly its extension towards p1$^{\prime}$, but still on the negative magnetic-field polarity), p1 and n0S, are the places where field lines in the fan of null N3 are anchored (compare Figure~\ref{qsls-N1-N3}(c) to 
Figure~\ref{null1-3}(b)). Although we cannot clearly isolate the emission due to reconnection in this null point, mainly because of its local nature, we think that it is present at different times during the series of events discussed in 
Section~\ref{S_Fil_Eruption}; notice, in particular, the brightest region to the SE in Figure~\ref{qsls-N2}(b).     

Another aspect, which is evident from the extension of the elongated brightening on n0S to the West, is that filament F1 expansion also forced reconnection at the QSLs associated to null N2 at this earlier time period; {\it i.e.} the shape of this portion of the bright band is similar to that of the QSL on n0S shown in Figure~\ref{qsls-N2}, although it is less extended and located eastwards.  

Once filament F1 erupted, AIA observed two ribbons at both sides of the PIL where the filament was. The evolution of flare FL1 can be explained by the ``standard'' (or CSHKP) model for eruptive flares by \citet{Carmichael64}, \citet{Sturrock66}, \citet{Hirayama74}, and \citet{Kopp76}. Although the intensity of the emission does not let us see the two ribbons clearly around flare maximum, a ``post-flare'' arcade appeared visible ($\approx$ 07:35 UT) on both sides of the PIL when the emission started decaying. 
        
\subsection{QSLs and the Flare FL2}
\label{S_QSLs_FL2}

\begin{landscape}
\begin{figure} 
\centering
\includegraphics[width=1.25\textwidth]{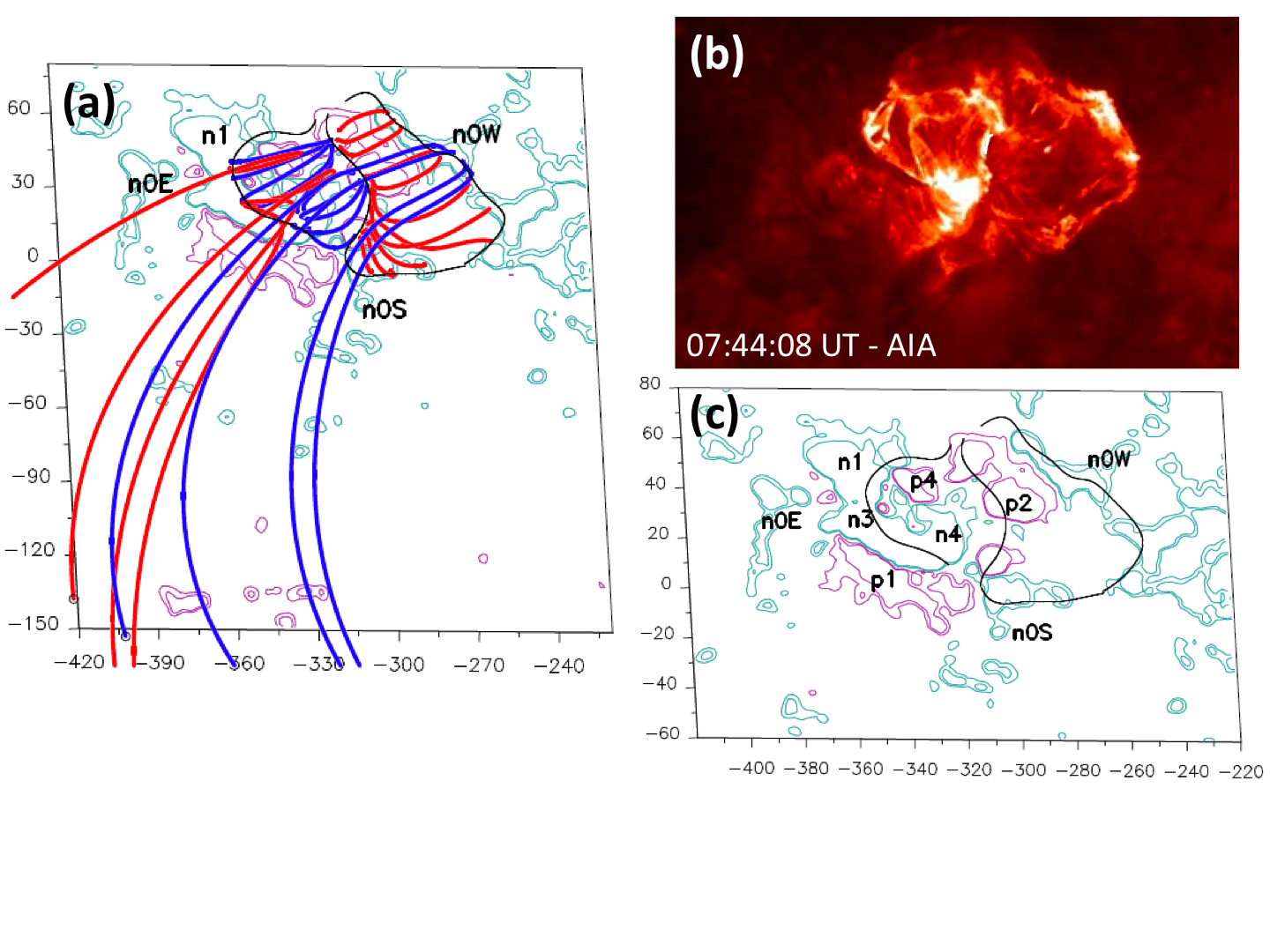}
\vspace*{-2.cm}
\caption{Comparison of QSLs and ribbons of flare FL2.
  (a) Photospheric trace of QSLs in AR 11123 computed on the polarities associated with null point N2 and field lines originating in the QSLs. Blue and red field lines indicate pre-reconnected and reconnected loops. The field lines that leave the drawn box correspond to large-scale loops connecting to the westernmost positive polarity in AR 11121
(Figure~\ref{null2}(a)).
 (b) An AIA 304 \AA\ image in which the roundish ribbon and central bright bar are present on polarities n0W, p2 and n1--n3--n2--n4. A set of post-flare loops from flare FL1 are seen to the East of the brightest region at the South East in this image; the presence of this bright region could indicate that reconnection was also at work in the localised null N3 and associated separatrices/QSls. The size of this panel is $210''$ ($120''$) in the horizontal (vertical) direction. Its centre in heliographic coordinates is located at [-175$''$,-430$''$]; Sun's centre is at [0$''$,0$''$].
  (c) Photospheric QSL traces and magnetogram.  The field of view is the same as that of panel (b). There is a close resemblance between the shape of QSLs and the AIA ribbons, indicating that they may originate due to reconnection at N2 and associated separatrices/QSLs.
  Both in (a) and (c), all axes are in Mm and the isocontours of the field correspond to $\pm$ 50, $\pm$ 100 G in
continuous magenta (blue) style for the positive (negative) values.
}
\label{qsls-N2}
\end{figure}
\end{landscape}

The comparison of the photospheric trace of QSLs (black continuous lines) located along the polarities associated to null N2 and flare FL2 emission is illustrated in Figure~\ref{qsls-N2}. As happens for the QSLs associated to null N1, the value of $Q$ is extremely high, being $\ge$ 10$^{10}$ for the traces shown in Figures~\ref{qsls-N2}(a) and (c).
The QSL traces closely resemble the shape of the roundish bright ribbon of flare FL2 and its central bar (compare panels (b) and (c) in Figure~\ref{qsls-N2}). A set of field lines computed starting integration at both sides of the QSLs has been included in Figure~\ref{qsls-N2}(a). Those computed from the outer border of the roundish QSL connect to an elongated region in the preceding western positive polarity of AR 11121 (see Figure~\ref{null2}), which we will call pW.
These results indicate that flare FL2 results from magnetic reconnection at N2 and its associated separatrices/QSls.

\begin{landscape}
\begin{figure} 
\centering
\includegraphics[width=1.35\textwidth]{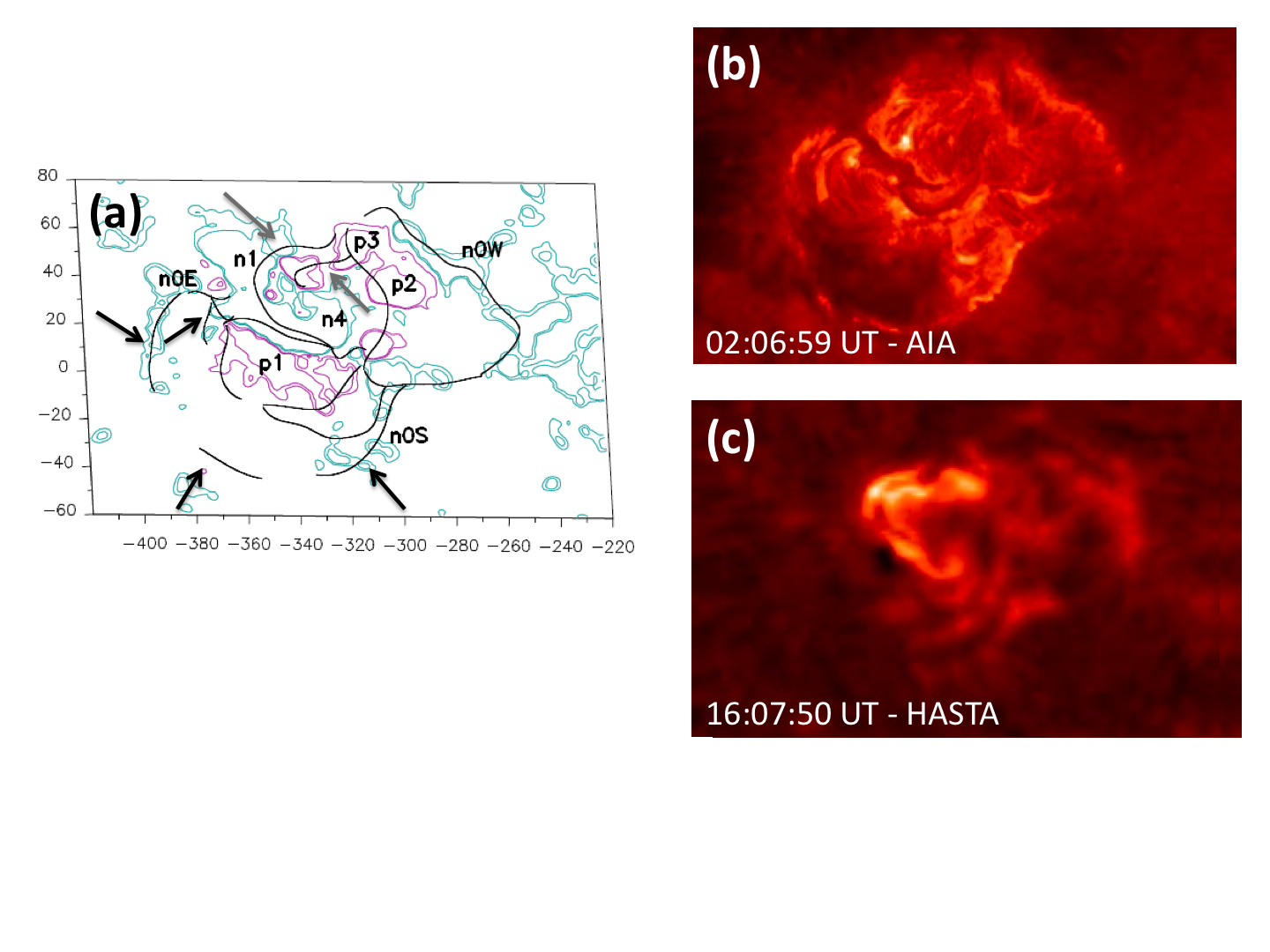}
\vspace*{-2.5cm}
\caption{(a) QSL trace related to the flares FL1 and FL2 (see Figures~\ref{qsls-N1-N3} and~\ref{qsls-N2}) plus those associated with a previous and later flare. Axes are in Mm and the isocontours of the field correspond to $\pm$ 50 and $\pm$ 100 G in continuous magenta (blue) style for the positive (negative) values.
To guide the eye the black (grey) arrows point to the QSLs along which the flare emision of the event in panel (b) ((c)) lay.  
  (b) Flare ribbons in 304 \AA\ for the flare with maximum at 02:14 UT in GOES curve (see Figure~\ref{GOES}). The size of this panel is $210''$ ($132''$) in the horizontal (vertical) direction. Its centre in heliographic coordinates is located at [-220$''$,-430$''$]; Sun's centre is at [0$''$,0$''$].
  (c) The two H$\alpha$ ribbons for the flare with maximum at 16:16 UT in GOES curve (see Figure~\ref{GOES}). The size of this panel is the same as that of panel (b) and its centre in heliographic coordinates is located at [-90$''$,-430$''$]; Sun's centre is at [0$''$,0$''$].
Panels (b) and (c) cover the same area as panel (a).  The QSL trace follows approximately the shape of both flare ribbons implying that magnetic reconnection at QSLs can also explain the origin of these events. 
} 
\label{qsls-extra}
\end{figure}
\end{landscape}

A section of filament F1 lay along the PIL between polarities n4 and p1$^{\prime}$--p2  (see Figure~\figAIA (a) and the set of orange field lines in Figure~\ref{nlfff}(a)). Although this section did not erupt, it is highly probable that it evolved and brighten during the eruption of the southern section forcing reconnection at null N2 and associated separatrices/QSLs, while flare FL1 was already underway. The set of field lines lying above this filament section would then reconnect to the large-scale lines connecting n0W to pW. Both sets have been drawn in blue in Figure~\ref{qsls-N2}(a). The resulting reconnected field lines would connect n0W to p1$^{\prime}$-p2 and the chain of negative polarities n1--n4 to pW. These field lines are shown in red in Figure~\ref{qsls-N2}(a). As for null N1, this reconnection process is expected to start at null N2 neighbourhood and its separatrices, where currents are the strongest, and evolve to involve the corresponding QSLs \cite{Masson09}. Notice that, at the time of the image shown in Figure~\ref{qsls-N2}(b), the brightest part of flare FL2 ribbons agrees with the location of the footpoints of field lines drawn in the fan of null N2 in Figure~\ref{null2}(b). Finally, as flare FL2 evolved, we expect to see flare ribbons at the QSL along the chain of negative polarities n1--n4, the QSL on p1$^{\prime}$--p2 and that on n0W, giving the roundish feature and the bright central bar. We should also observe an enhancement on pW but, as in the case of the brightening on pSE, this is very unlikely for the same reasons. 

\subsection{Other Flaring Episodes in AR 11123}
\label{S_QSLs_Other}

When computing the QSLs associated with nulls N1 and N2, we also find other traces at different locations from those corresponding to the events discussed in the previous section. 
Figure~\ref{qsls-extra}(a) illustrates the locations of the full set of QSLs, those pointed with black (grey) arrows follow approximately the shape of the ribbons of the flare C2.9 (C4.3) starting at around 01:58 UT (15:53 UT), which is shown in Figure~\ref{qsls-extra}(b) 
(Figure~\ref{qsls-extra}(c)). For these traces the value of $Q$ is 
much lower (between 10$^3$ and 10$^4$). 

It is not our aim to discuss the C2.9 and C4.3 flares in detail in this article, we only want to stress two aspects. First, these events can also be explained by reconnection at QSLs. Second, although the magnetogram used as boundary condition to compute the locations of QSLs is closer in time to the events described in Section~\ref{S_Fil_Eruption}, we also find the QSLs associated with these two flares. This is so because, as stated before, the location of QSLs is strongly determined by the distribution of the polarities at the photosphere. In this respect, we remark that by early 11 November all of the polarities in AR 11123 had already emerged and we neither measure a further magnetic-flux increase (see Figure~\ref{flux-evol}) nor observe relevant photospheric polarity displacements along that day.     

\section{Summary and Discussion}
\label{S_Interpretation}

A complex, formed by the decaying AR 11121 and the new emerging AR 11123, produced ten flares on 
11 November 2010. The new AR, which was born in the negative environment of the following polarity of AR 11121, was composed by a series of bipoles that emerged violently in less than one day. The activity in the complex occurred mainly in this new AR at recurrent locations and was related to the activation and, sometimes, eruption of short and low-lying AR filaments (F1, F2 and F3). 
To interpret the observations, obtained by instruments onboard SDO (AIA and HMI) and from the ground (THEMIS and HASTA), and understand the origin of several flares and a filament eruption, we have modelled the magnetic field of the complex in the LFFF and NLFFF approaches.    

The presence of coronal magnetic null points is an indication of a complex topology, {\it i.e.} a configuration where magnetic reconnection can occur (see references in Section~\ref{S_Introduction}). We find three coronal magnetic null points in AR 11123 at similar locations and with similar characteristics in the LFFF and the NLFFF approaches: two positive nulls (N1 and N2) and a negative null (N3). These similarities give support to their existence.  

The understanding of the role of brightenings preceding filament eruptions and/or flares is crucial to learn how the eruptive configuration is built up and driven towards an instability. Preflare events have not been analysed frequently \citep[{\it e.g.}][]{Farnik96,Williams05,Mandrini06} because these precursors are low-energy events which do not stand out against other background emissions. 

We find that magnetic reconnection at the southernmost null point [N1] plays a key role during the eruption of filament F1. This null point has an asymmetric fan structure since, both in the LFFF and the NLFFF extrapolations, one of the fan eigenvalues has a larger amplitude, by a factor between $\approx$ 7 and $\approx$ 9, than the other one. This implies a very asymmetric magnetic configuration around N1 with field lines concentrating around the eigenvector with the largest eigenvalue. We observe a compact brightening approximately at the location of the fan field-line footpoints to the North and a very elongated brightening to the South. These are precursors to flare FL1. No brightening is discernible either at the location of the lower spine-related field-line footpoints, since it could be masked by the expanding filament F1, or at the upper spine-related field-line footpoints in the preceding positive polarity of AR 11121, since energy deposited in the very long reconnected field lines could be dissipated before reaching the chromosphere. Furthermore, the evolution of the elongated AIA brightening and the different steps during the expansion of filament F1 are not conclusive in relation to the scenario that forces its eruption, which is followed by the two-ribbon flare FL1. We cannot discriminate between a break-out and a tether-cutting process. However, it is clear that filament F1 eruption and null-point reconnection ``interact'' in a positive feedback way. The filament eruption drives null-point reconnection, while the latter facilitates the former by reducing the magnetic tension.

 Next, our topological analysis indicates that magnetic reconnection at the northernmost null point [N2] is fundamental to understand the location of the ribbons during flare FL2. As stated in Section~\ref{S_QSLs_FL2}, it is highly plausible that the destabilisation of filament F1 drives reconnection at null N2 giving rise to the confined flare FL2. The fan of null N2 is highly asymmetric in the LFFF model, while this asymmetry is much less in the NLFFF extrapolation. This is the main difference we have found between the two modelling approaches in relation to magnetic null points. On one hand, flare FL2 presents a roundish ribbon, as expected, and the emission intensity is higher at the locations where the fan field-line footpoints cluster due to the difference between the amplitude of the fan eigenvalues. On the other hand, flare emission from inside the roundish ribbon, where the inner spine footpoint is located, is not compact; it extends forming a bright band, then the null-point topology is not sufficient to understand the observed ribbons. 
 
The extension of the observed precursor brightenings of flare FL1 and the intensity distribution of flare FL2 ribbons are both explained by the location of QSLs in AR 11123. The traces of QSLs agree closely with all precursor and flare FL2 brightenings and, as in \citet{Masson09} and \citet{Reid12}, the magnetic null-point separatrices are embedded in more extended QSLs. Furthermore, the results of our more complete topological analysis are also useful to understand two other events, one previous to and one following the ones that we analyse in detail. Reconnection proceeding within these more general topological structures was found in MHD simulations \cite{Aulanier05,Aulanier06,Masson09}. QSLs can explain more diverse events than the magnetic topology limited to separatrices can do (Section~\ref{S_Introduction}). This is further shown in the events studied since, computing only magnetic null points and their associated separatrices, it would not have been possible to understand the full extension of the observed events.

{\em Acknowledgements}. We thank the Flux Emergence team led by K. Galsgaard and F. Zuccarello for fruitful discussions during the ISSI workshops in Bern, Switzerland. Data used in this article are courtesy of NASA/SDO and the AIA and HMI science teams. THEMIS is a French telescope installed in the Canary Islands and funded by CNRS/INSU. We thank the team of THEMIS for these observations and particularly V. Bommier who was the PI during the campaign and ran the {\sf UNNOFIT} inversion code. This study uses data obtained at OAFA (El Leoncito, San Juan, Argentina) in the framework of the German--Argentinean HASTA/MICA Project, a collaboration of MPE, IAFE, OAFA and MPAe. CHM and GDC acknowledge financial support from the Argentinean grants PICT 2007-1790 (ANPCyT), UBACyT 20020100100733 and PIP 2009-100766 (CONICET). CHM and GDC are members of the Carrera del Investigador Cient\'\i fico (CONICET). YG was supported by the National Natural Science Foundation of China (NSFC) under
grant numbers 11203014, 10933003 and by a grant from the 973 project 2011CB811402.

\bibliographystyle{elsart-harv}

\bibliography{mandrini_Nov10}

\IfFileExists{\jobname.bbl}{} {\typeout{}
\typeout{****************************************************}
\typeout{****************************************************}
\typeout{** Please run "bibtex \jobname" to obtain} \typeout{**
the bibliography and then re-run LaTeX} \typeout{** twice to fix
the references !}
\typeout{****************************************************}
\typeout{****************************************************}
\typeout{}}

\end{document}